\documentclass{article}
\pdfoutput=1 
\usepackage{arxiv}
\usepackage[round]{natbib}

\usepackage[utf8]{inputenc} 
\usepackage[T1]{fontenc}    
\usepackage{url}            
\usepackage{booktabs}       
\usepackage{nicefrac}       
\usepackage{microtype}      
\usepackage{natbib}
\usepackage{doi}

\usepackage{lineno,hyperref}
\modulolinenumbers[5]
\usepackage{amsmath}
\usepackage{amsfonts}
\usepackage{comment}
\DeclareMathOperator*{\argmax}{arg\,max}

\usepackage{algorithm}
\usepackage[noend]{algpseudocode}
\usepackage{graphicx}
\usepackage{adjustbox}
\usepackage{subcaption}
\usepackage{placeins}
\usepackage{colortbl}
\usepackage{diagbox}
\usepackage{multirow}
\usepackage{textcomp}
\usepackage[british]{babel}
\usepackage{csquotes}

\title{Act to Reason: A Dynamic Game Theoretical Model of Driving}

\author{ Cevahir~Köprülü \\
	Department of Electrical and Electronics Engineering\\
	Bilkent University\\
	Ankara, Turkey 06800 \\
	\texttt{cevahir.koprulu@ug.bilkent.edu.tr} \\
	\And
	Yıldıray~Yıldız\thanks{Corresponding author} \\
	Department of Mechanical Engineering\\
	Bilkent University\\
	Ankara, Turkey 06800 \\
	\texttt{yyildiz@bilkent.edu.tr} \\
}

\date{}

\renewcommand{\headeright}{}
\renewcommand{\undertitle}{}

\begin{document}
\maketitle

\begin{abstract}
The focus of this paper is to propose a driver model that incorporates human reasoning levels as actions during interactions with other drivers. Different from earlier work using game theoretical human reasoning levels, we propose a dynamic approach, where the actions are the levels themselves, instead of conventional driving actions such as accelerating or braking. This results in a dynamic behavior, where the agent adapts to its environment by exploiting different behavior models as available moves to choose from, depending on the requirements of the traffic situation. The bounded rationality assumption is preserved since the  selectable strategies are designed by adhering to the fact that humans are cognitively limited in their understanding and decision making. Using a highway merging scenario, it is demonstrated that the proposed dynamic approach produces more realistic outcomes compared to the conventional method that employs fixed human reasoning levels. 
\end{abstract}



\section{Introduction}

Modeling human driver behavior in complicated traffic scenarios paves the way for designing autonomous driving algorithms that incorporate human behavior. Complementary to behavior integration, building realistic simulators to test and verify autonomous vehicles (AVs) is imperative. Currently, the pace of AV development is hindered by demanding requirements such as millions of miles of testing to ensure safe deployment \citep{KALRA2016182}. High-fidelity simulators can boost the development phase by constructing an environment that is comprised of human-like drivers \citep{doi:10.1007/978-3-642-00602-9_28, doi:10.1109/9.664155, KAMALI201788}. In addition, human-driver models may induce the development of safer algorithms. Deploying control systems that behave similar to humans enables a familiar experience for the passengers and a recognizable pattern for surrounding drivers to interact with. Therefore, designing models that display human-like driving is vital. 

Various driver models employing a range of methods are proposed in the literature. Examples of inverse reinforcement learning based approaches, where the agents' preferences are learned from demonstrations can be found in the studies conducted by \cite{7139555,7995721,sadigh2018planning,8569453}. In addition, an example of direct reinforcement learning method can be found in the work of \cite{trans_rpc_DRL_car_following}. Data-driven machine learning methods are commonly proposed to model driver behavior, as well \citep{trans_rpc_data_driven_LC,4290188,6670061,trans_rpc_random_forest,trans_rpc_gradient_boosting_logit,trans_rpc_LC_speech_recognition,trans_rpc_estim_style,7535533,li2019coordination,7795678,trans_rpc_work_zone_merging,7995935,8317887,trans_rpc_driving_style}. Different from machine learning methods, studies that utilize game theoretical approaches investigate strategic decision making. Studies that employs these approaches are presented by \cite{trans_rpc_GT_based_LC,trans_rpc_GT_social_interaction,yoo2013stackelberg,ALI2019220,schwarting2019social} and \cite{8794007}. Apart from machine learning and game theory based concepts, there exist studies following a control theory direction \citep{trans_rpc_fuzzy,8057588}.

In recent years, a concept based on \enquote*{Semi Network Form Games}  \citep{Lee2012, lee_wolpert_bono_backhaus_bent_tracey_2013} emerged that combines reinforcement learning and game theory, to model human operators both in the aerospace \citep{doi:10.2514/6.2012-4487, doi:10.2514/1.G000176, doi:10.2514/1.G000426} and automotive domains \citep{7993050, ALBABA20191, 8814955, 7798354, 7525162, 9261970,8917314}.
One of the main components of this method is the level-\textit{k} game theoretical approach \citep{STAHL1995218,costa2009comparing,camerer2011behavioral}, which assigns different levels of reasoning to intelligent agents. The lowest level, level-0 is considered to be non-strategic since it acts based on predetermined rules without considering other agents' possible moves. A level-1 agent, on the other hand, provides a best response assuming that the others are level-0 agents. Similarly, a level-\textit{k} agent, where k is an integer, acts to maximize its rewards based on its belief that the environment agents have level-($k-1$) reasoning. In all of these earlier studies, it is assumed that agents' reasoning levels remain the same, throughout their interactions with each other. Although this assumption may be valid for the initial stages of interaction, it falls short of modeling adaptive behavior. For example, drivers can and do adapt to the driving styles of the other drivers, which is not possible to model with the conventional fixed level-\textit{k} approach. In this paper, we fill this gap in the literature and introduce a \enquote*{dynamic level-\textit{k}} approach where the human drivers may have varying levels of reasoning.  We achieve this goal by transforming the action space from direct driving actions, such as \enquote*{accelerate}, \enquote*{break}, or \enquote*{change lane}, to reasoning levels themselves. This is made possible by using a two-step reinforcement learning approach: In the first step, conventional level-\textit{k} models, such as level-1, level-2 and so on are trained. In the second step, the models are trained again using reinforcement learning but this time they use the reasoning levels, which are obtained from the previous step, as their possible \enquote*{actions}.

There are other studies which also incorporate a dynamic level-\textit{k} reasoning in agent training. \cite{gt_unsig_inter_Li_2} utilized receding-horizon optimal control to determine agent's actions, where the opponent's reasoning level is deduced from a probability distribution, which is updated during the interaction. \cite{adaptive_gt_roundabouts_Li_1} developed a controller which directly predicts opponent's level based on its belief function, and acts accordingly. Both of these studies model the interactions of two agents at a time, which makes them nontrivial to extend for modeling larger traffic scenarios. Our study distinguishes itself by designing agents that dynamically select a reasoning level given only their partial observation of the environment, instead of employing a belief function. This allows our method to naturally model crowded traffic scenarios. \cite{doi:10.1287/mnsc.1120.1645} also proposed a dynamic level-\textit{k} method, where, similar to the ones discussed above, the opponent's levels are predicted using belief functions, and therefore it may not be suitable for modeling crowded multi-player games. \cite{doi:10.1287/mnsc.2020.3595} extended their previous work for n-player games by proposing an iterative adaptation mechanism, where the ego agent assumes that all of its opponents have the same reasoning level. This may be restricting for heterogeneous traffic models.

Specifically, the contributions of this study are the following:

\begin{enumerate}
    \item We introduce a dynamic level-\textit{k} model, which incorporates a real-time selection of driver behavior, based on the traffic scenario. It is noted that individual level-\textit{k} models have reasonable match with real traffic data as previous studies show \citep{ALBABA20191}. In this regard, we preserve bounded rationality by selecting agents' actions only from a  level-\textit{k} strategy set. 
    \item The proposed method allows a single model for driver behavior on both the main road and the ramp in the highway merging setting. In general, earlier studies only consider one or the other scenario due to the complicatedness of the problem at hand. 
    \item In comparison to other methods, large traffic scenarios with multiple vehicles can naturally be modeled, which suits better for the complexity of real-life conditions.
\end{enumerate}

This paper is organized as follows. In Section \ref{section:background}, level-\textit{k} game theory and Deep Q-learning are briefly described. In Section \ref{section:methods}, proposed dynamic level-\textit{k} model is explained. In Section \ref{section:scenario}, the construction of highway merging environment based on NGSIM I-80 dataset \citep{NGSIM_I80} is demonstrated along with the observation and action spaces, vehicle model and the reward function. In Section \ref{section:training_and_simulation}, training configuration, implementation details, level-0 policy, and the results of training and simulation are provided. Finally, a conclusion is given in Section \ref{section:conclusion}.

\vspace{-.1cm}

\section{Background}
\label{section:background}

In this section, we provide the main components of the proposed method. Namely, level-\textit{k} reasoning, Deep Q-learning (DQN) and the synergistic employment of level-\textit{k} reasoning and DQN are explained. For brevity, we provide only a brief summary, the details of which can be found in the work of \cite{albaba2020driver}.

\subsection{Level-\textit{k} Game Theory}

Level-\textit{k} game theory is a non-equilibrium hierarchical game theoretic concept that models strategic decision making process of humans \citep{10.1162/0033553041502225, STAHL1995218}. In a game, where N players are interacting, a level-\textit{k} strategy maximizes the utility of a player given that the other players are making decisions based on the level-($k-1$) strategy. Therefore, in a game with players $p_i$, $i \in \{0,1,2,\cdots,N-1\}$, we can define the level-\textit{k} policy as

\begin{equation}
    \pi^k = \argmax_{\pi}u_i(\pi | \pi_{p_1},\pi_{p_1},\cdots,\pi_{p_{N-1}}),
\end{equation}
where $\pi_{p_i}=\pi^{k-1}$ $\forall i \in \{1,2,\cdots,N-1\}$. 

In practice, a level-0 player is determined as a non-strategic agent whose actions are selected based on a set of rules that may or may not be stochastic. Then, a level-1 player that provides the best responses to the level-0 decision makers is formed. Similarly, a level-2 agent, who provides the best responses to a level-1 player, is created. Iteratively, this process continues until a final level-$K$, $K \in \mathbb{Z^+}$, strategy is obtained.

\subsection{Merging Deep Q-Learning with Level-\textit{k} Reasoning}

In early implementations, where reinforcement learning and game theory is merged \citep{7993050, ALBABA20191, 8814955, 7798354, 7525162, 9261970}, a tabular RL method was used. Due to the requirement of an enlarged observation space to obtain higher-fidelity driver models, recently a Deep Q-Learning approach is introduced \citep{albaba2020driver}.
Deep Q-learning, \citep{mnih2013playing, mnih2015human}, utilizes a neural network of fully-connected or convolutional layers and maps an observed state $s_t$ at time $t$ from the state space $S$ to an action value function $Q(s_t,a_t)$, where $a_t \in A$ is the action taken at time $t$. DQN algorithm achieves this goal using Experience Replay and Boltzmann Exploration \citep{ravichandiran2018hands}. 

In the context of obtaining high fidelity driver models, DQN is used to obtain the level-\textit{k} driver behaviour by training the ego vehicle's policy in an environment where all other drivers are assigned a level-($k-1$) policy. The details of the algorithm merging DQN and level-\textit{k} reasoning is explained by \cite{albaba2020driver}.

\section{Methods}
\label{section:methods}

In this section, dynamic level-\textit{k} approach is explained in detail. 

\subsection{Dynamic Level-\textit{k} Model}

Level-\textit{k} strategy is built on a condition that the ego player is interacting with agents, who have level-($k-1$) policies. Such an assumption brings about a problem when the agent is in an environment consisting of agents with policies that are different than level-($k-1$). In that case, the agent may make incorrect assumptions about others' intentions and act in a way that can be detrimental to itself and others. As a countermeasure, we train the agent to select a policy among available level-\textit{k} policies at each time-step in order to maximize its utility in mixed traffic.

\begin{figure}[h!]
    \centering
    \includegraphics[scale=.5]{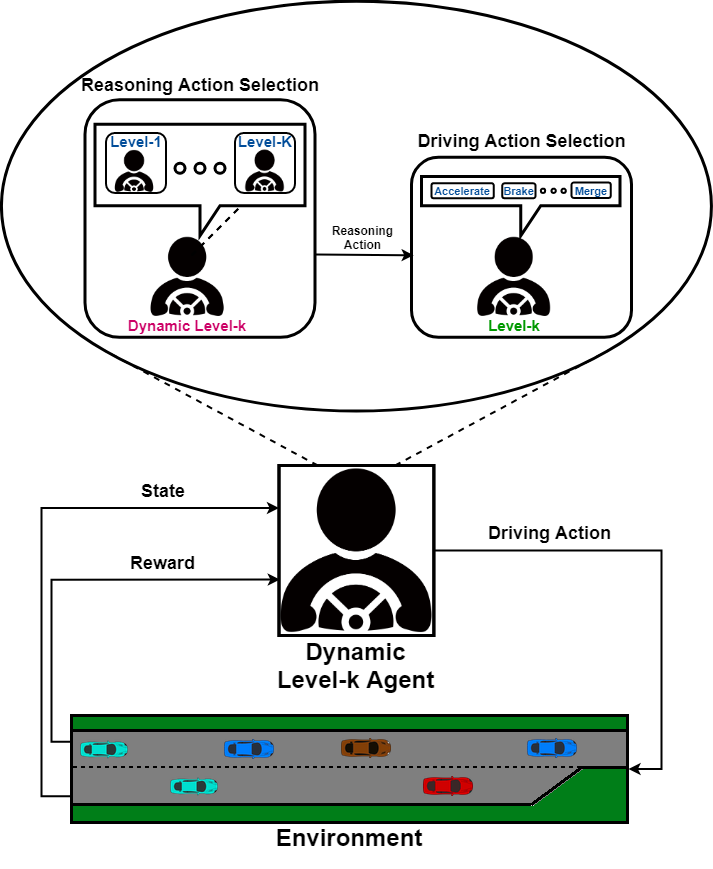}
    \caption{Dynamic Level-\textit{k} Model}
    \label{fig:graphical_abstract}
\end{figure}

We start by defining a set of ($K$+$1$)-many level-\textit{k} policies as
\begin{equation}
    \Pi = \{ \pi^0, \pi^1, \pi^2, \cdots , \pi^{K} \}.
\end{equation}
This set determines the available level-\textit{k} policies, where each policy, with the exception of level-0 policy, which is pre-determined, is trained in an environment where the rest of the drivers are taking actions based on level-($k-1$) policies. To train the ego agent with a dynamic level-\textit{k} policy, we place it in an environment with $N$ many cars, including the agent. From the ego's perspective, the environment is comprised of ($N-1$)-fold Cartesian product of policies, represented as

\begin{equation}
    \Lambda_N = \Pi \times \Pi \times \cdots \times \Pi.
\end{equation}
An element $\lambda$ of the set $\Lambda_N$ is an ordered $N-1$ tuple, expressed as

\begin{equation}
    \lambda = (\pi_{p_1},\pi_{p_2}, \cdots, \pi_{p_{N-1}}) \in \Lambda_N,
\end{equation}
where $\pi_{p_i}$ is the policy of player $i$, $i \in \{1,2, \cdots N-1\}$. In this manner, the dynamic level-\textit{k} policy can be defined as

\begin{equation}
    \pi^{dyn} = \argmax_{\pi}u(\pi|\Lambda_N),
    \label{eq:dynamic_levelk}
\end{equation}
where u represents utility. The dynamic level-\textit{k} policy defined in (\ref{eq:dynamic_levelk}) is the policy that maximizes the utility of an agent placed in an environment consisting of ($N-1$)-fold Cartesian product of $K$ different level-\textit{k} policies. In the reinforcement learning setting, given an observation, the dynamic policy selects a level-\textit{k} policy among available level-\textit{k} policies, and then a direct driving action (breaking, acceleration, etc) is sampled using this selected policy. The working principle of the dynamic level-\textit{k} model is illustrated in Fig. \ref{fig:graphical_abstract}. 

\FloatBarrier

\noindent The algorithm for the training process  is given in Algorithm \ref{alg::dynamic}.

\noindent \textbf{Remark:} In the proposed approach, the selection of levels as actions are not based on the history of environment vehicles, which would be prohibitive for modeling crowded scenarios. Instead, the behavior selection is conducted using a policy, which provides a direct stochastic map between immediate observations to actions.  

\begin{algorithm}[H]
\caption{Training Dynamic Level-\textit{k} Decision Maker with DQN}
    \begin{algorithmic}[1]
        \State Initialize environment with $N$ vehicles
        \State Load level-\textit{k} policies $\pi^k \in \Pi$, trained earlier,  to networks $QN^k$ for $k=1,2,\cdots,K$
        \State Initialize the primary network $QN^{dyn}$ via Xavier uniform initializer \citep{pmlr-v9-glorot10a}
        \State Initialize the target network $QN^{dyn}_{target}$ with primary network weights
        \State Initialize memory $M$ as a queue with maximum size $N_{initial}$ 
        \State Initialize Boltzmann Temperature $T$ 
        \For {$e=1$ to $E$}
            \State Generate a random $N-1$ tuple $\lambda \in \Lambda_N$ for episode $e$
            \For {$t=1$ to $L$}
                \State Pass $s_t$ through $QN^{dyn}$ to obtain $Q_{\pi^{dyn}}(s_t,\pi^m), \forall \pi^m \in \Pi - \{\pi^0\}$
                \State Sample $\pi^m_t$ among available policies via Boltzmann Sampling
                \State Pass $s_t$ through $QN^m$ to obtain $Q_{\pi^m_t}(s_t,a_j), \forall a_j \in A$
                \State Sample an action $a_t$ via Boltzmann Sampling
                \State Transition into next state $s_{t+1}$ by executing action $a_t$
                \State Observe reward $r_t$ 
                \State Store current experience $(s_t,\pi^m_t,r_t,s_{t+1})$ in $M$
                \If{$size(M) \geq N_{initial}$}
                    \State Randomly sample $P$-sized batch of experiences $(s_i,\pi^k_i,r_i,s_{i+1})$
                    \For {$i=1$ to $P$}
                        \If{$s_{t+1}$ is $terminal$}
                            \State $y_i = r_i$
                            \Else
                            \State $y_i = r_i + \gamma \max_{\pi'}QN^{dyn}_{target}(s_{i+1},\pi';\theta^{dyn}_{target})$
                            \EndIf
                        \State Calculate loss $(y_i -QN^{dyn}(s_i,\pi^k_i;\theta^{dyn}))^2$
                        \State Perform gradient descent using loss on primary weights $\theta^{dyn}$
                        \EndFor
                    \EndIf
                \State Every $U$ steps, set $\theta^{dyn}_{target} := \theta^{dyn}$
                \If{$s_{t+1}$ is $terminal$}
                    \State End episode $e$
                    \EndIf
                \EndFor
            \State Update Boltzmann temperature: $T = \max (T*c,1 )$, $c < 1$
            \EndFor
    \end{algorithmic}
    \label{alg::dynamic}
\end{algorithm}

\section{Scenario Construction Based On Traffic Data}
\label{section:scenario}

We implement the proposed dynamic level-\textit{k} approach in a highway merging scenario. To demonstrate the capabilities of the new method, we choose to create a realistic scenario that represents a real traffic environment. To achieve this, we employ NGSIM I-80 data set \citep{NGSIM_I80} to construct the road geometry (see Fig. \ref{fig:I80_area}) and to impose constraints on the driver models. 

\subsection{Highway Merging Setting}

Highway merging scenario primarily consists of 2 lanes, one of which is part of the main road and the other is the ramp. In comparison to highway driving, highway merging is a more challenging problem, whose participants are grouped into  two types: The first type attempts to merge from the ramp to the main road, and the second type commutes on the main road. Its exhausting nature is caused by mandatory lane changes and frequently varying velocities of the participants. As the traffic becomes more and more crowded, predicting where and how an interacting vehicle merges grows intensely troublesome. In this context, risky interactions occur more often than in a highway setting. Such a distinction is mainly brought about by the imminent trade-off between being safe and leaving the merging region as fast as possible. Essentially, how a driver handles this trade-off determines the preferred style.

\subsubsection{NGSIM I-80 Dataset Analysis}
\label{section:NGSIM_I80_Analysis}

\begin{figure}[t]
    \centering
    \includegraphics[scale=.7]{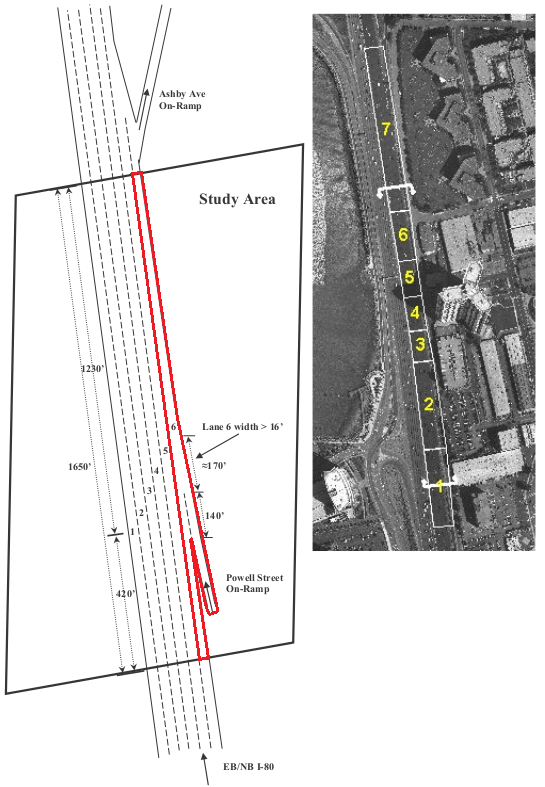}
    \caption{NGSIM I-80 Study Area Schematic \citep{NGSIM_I80}: Red region is the area of interest for this work.}
    \label{fig:I80_area}
\end{figure}

NGSIM I-80 dataset consists of vehicle trajectories from the northbound traffic in the San Francisco Bay area in Emeryville, CA, on April 13, 2005. The area of interest is around 500 meters in length and includes six highway lanes, and an on-ramp. Fig. \ref{fig:I80_area} shows a schematic of the area. 

\begin{figure}[h!]
\centering
    \begin{subfigure}[h!]{0.325\textwidth}
        \adjincludegraphics[width=\textwidth,Clip={0\width} {0\height} {0.075\width} {0\height}]{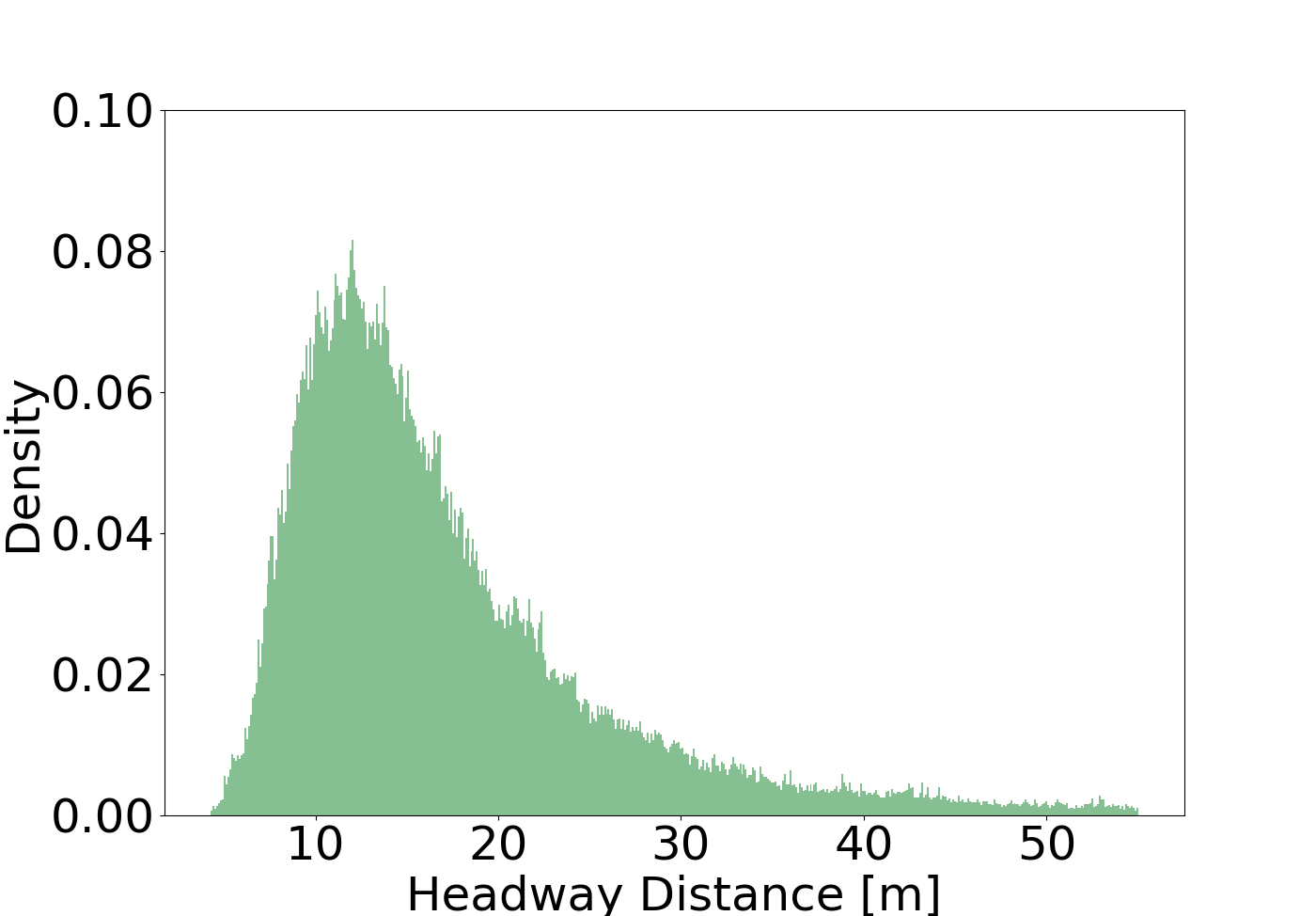}
        \caption{Main Road}
        \label{fig:headway_main}
    \end{subfigure}
    \hfill
    \begin{subfigure}[h!]{0.325\textwidth}
        \adjincludegraphics[width=\textwidth,Clip={0\width} {0\height} {0.075\width} {0\height}]{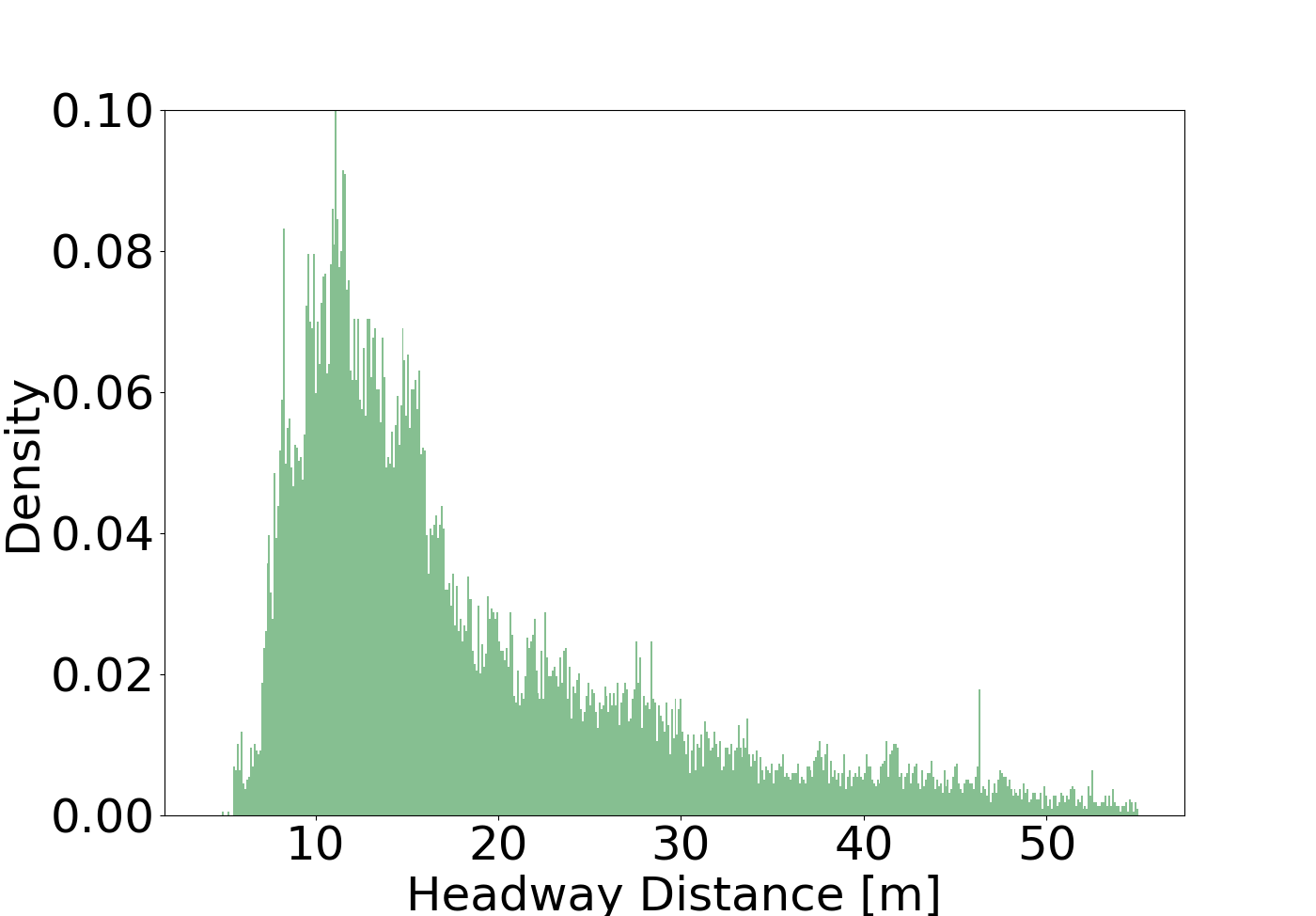}
        \caption{Ramp}
        \label{fig:headway_ramp}
    \end{subfigure}
    \hfill
    \begin{subfigure}[h!]{0.325\textwidth}
        \adjincludegraphics[width=\textwidth,Clip={0\width} {0\height} {0.075\width} {0\height}]{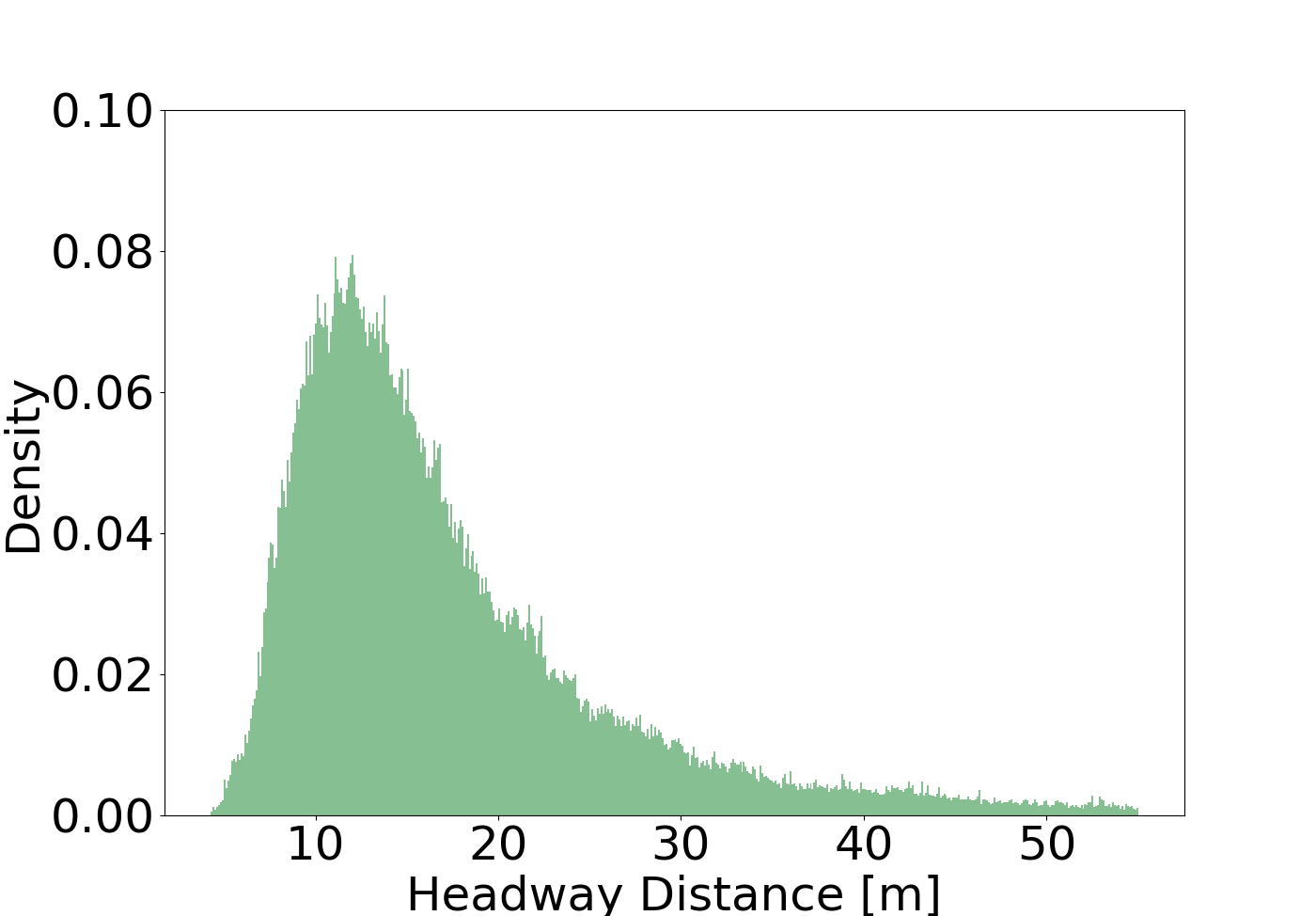}
        \caption{Both Lanes}
        \label{fig:headway_both}
    \end{subfigure}
\caption{NGSIM I-80 Headway Distance Distributions}
\label{fig:headway_dist}
\end{figure}

\begin{figure}[h!]
\centering
    \begin{subfigure}[h!]{0.325\textwidth}
        \adjincludegraphics[width=\textwidth,Clip={0\width} {0\height} {0.075\width} {0\height}]{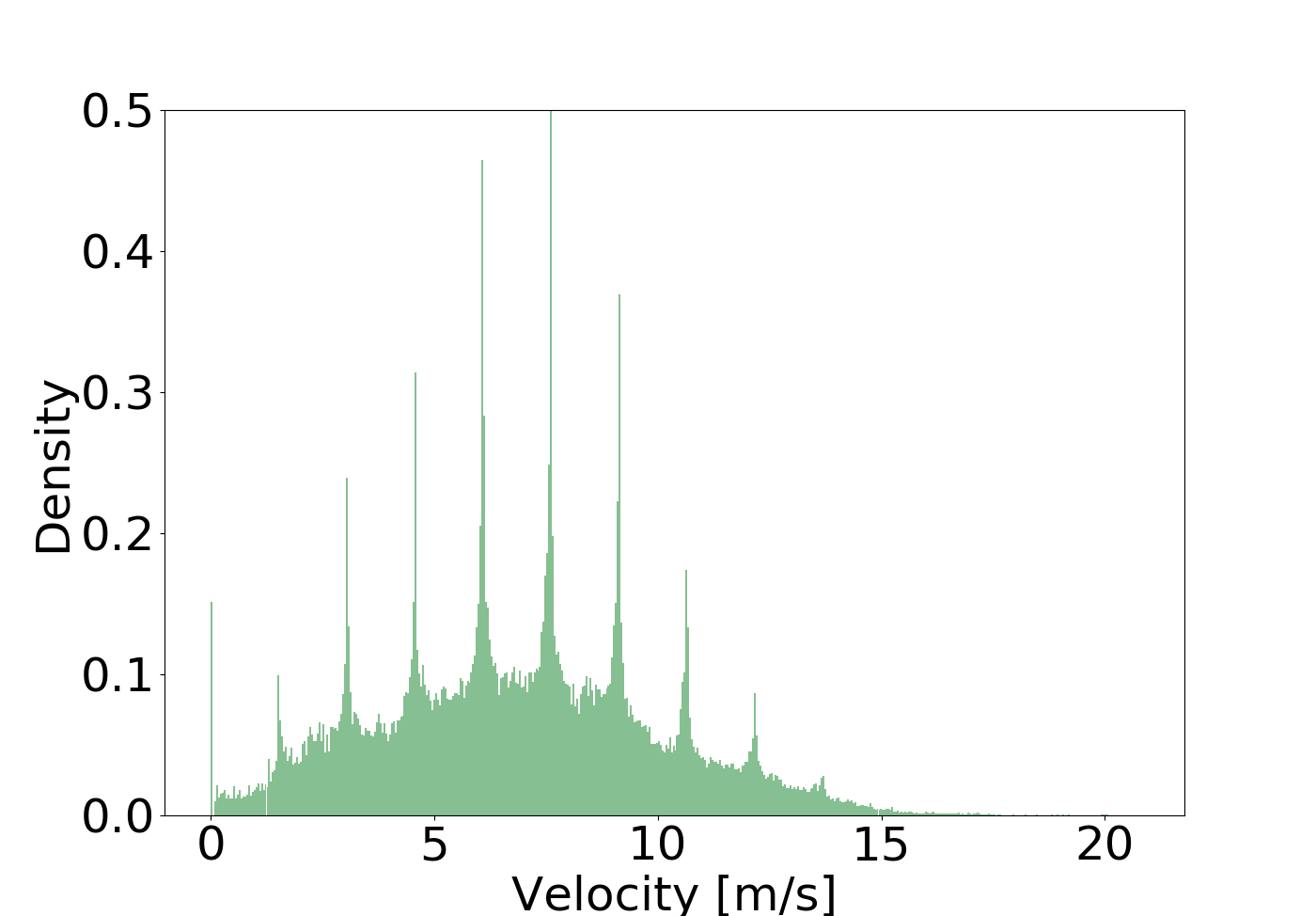}
        \caption{Main Road}
        \label{fig:vel_main}
    \end{subfigure}
    \hfill
    \begin{subfigure}[h!]{0.325\textwidth}
        \adjincludegraphics[width=\textwidth,Clip={0\width} {0\height} {0.075\width} {0\height}]{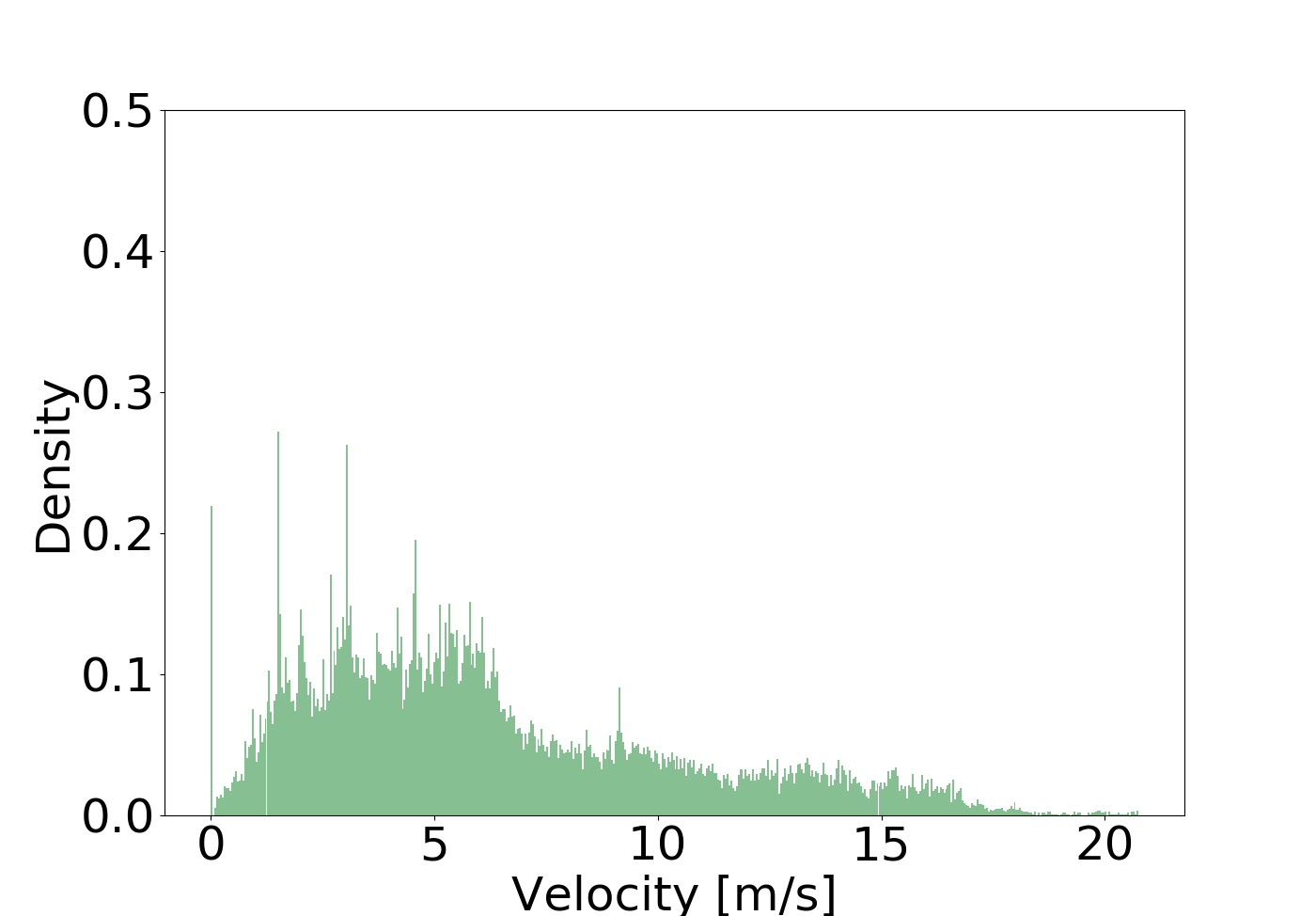}
        \caption{Ramp}
        \label{fig:vel_ramp}
    \end{subfigure}
    \hfill
    \begin{subfigure}[h!]{0.325\textwidth}
        \adjincludegraphics[width=\textwidth,Clip={0\width} {0\height} {0.075\width} {0\height}]{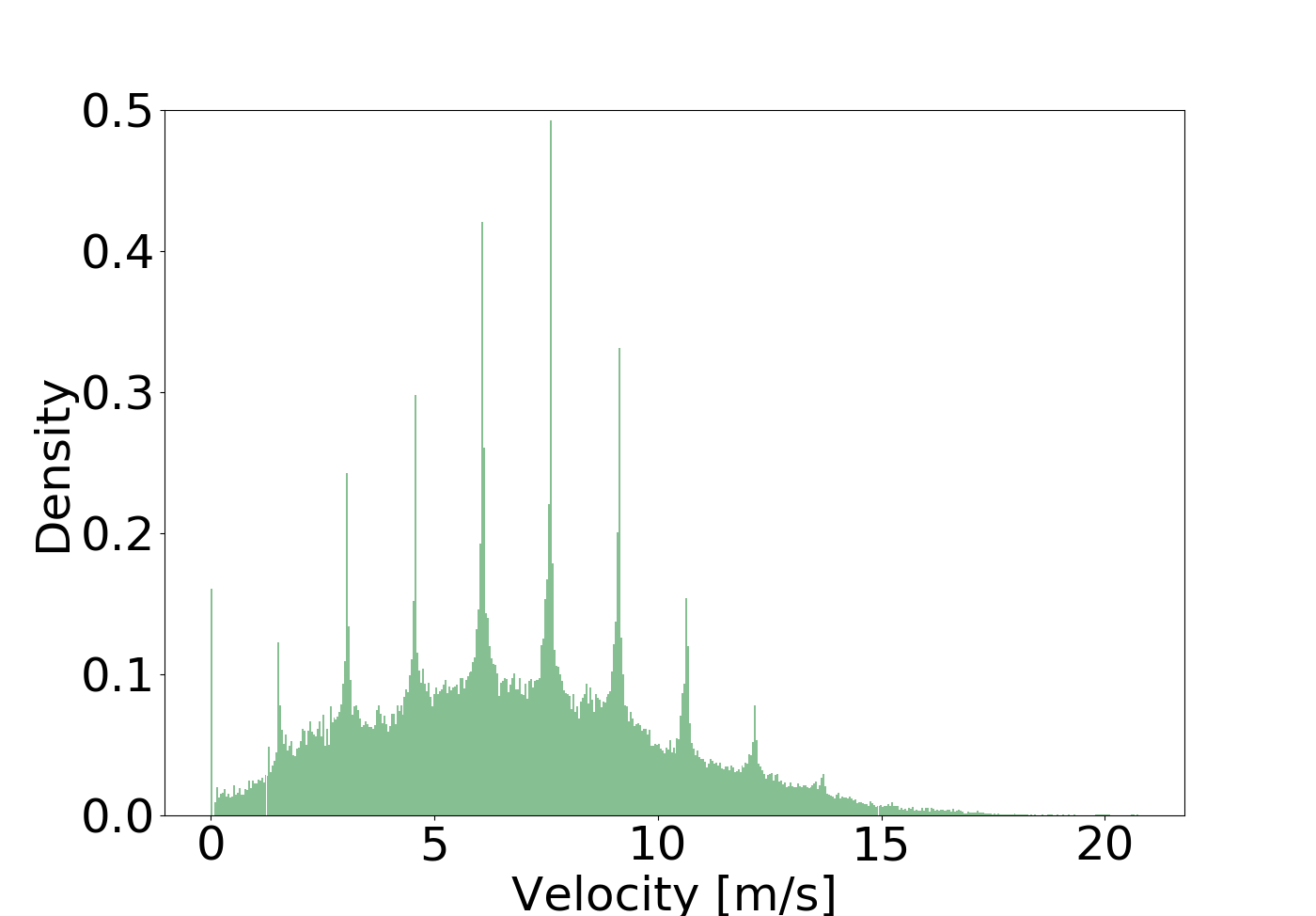}
        \caption{Both Lanes}
        \label{fig:vel_both}
    \end{subfigure}
\caption{NGSIM I-80 Velocity Distributions}
\label{fig:vel_dist}
\end{figure}

\begin{figure}[h!]
\centering
    \begin{subfigure}[h!]{0.325\textwidth}
        \adjincludegraphics[width=\textwidth,Clip={0\width} {0\height} {0.075\width} {0\height}]{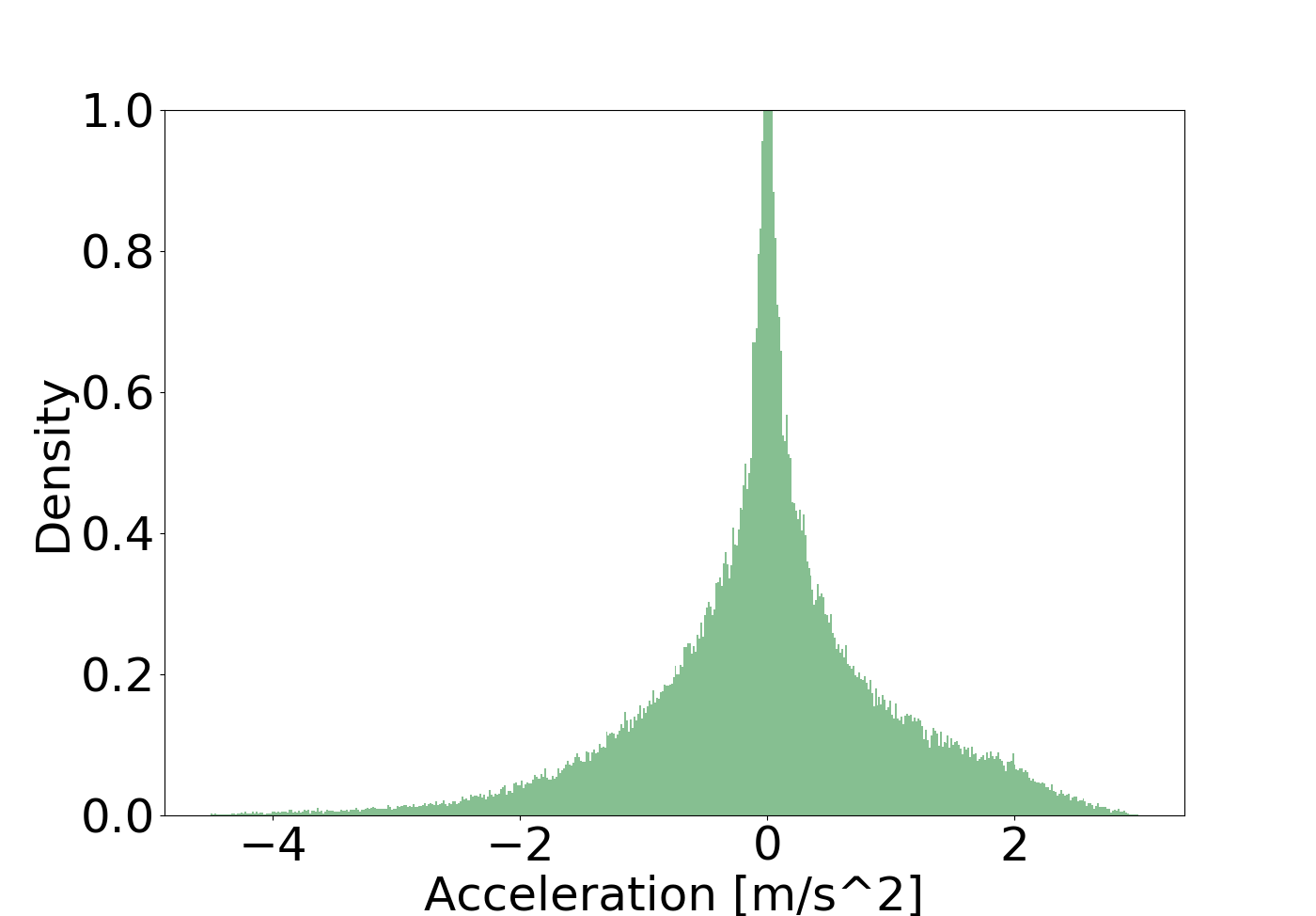}
        \caption{ Main Road}
        \label{fig:acc_main}
    \end{subfigure}
    \hfill
    \begin{subfigure}[h!]{0.325\textwidth}
        \adjincludegraphics[width=\textwidth,Clip={0\width} {0\height} {0.075\width} {0\height}]{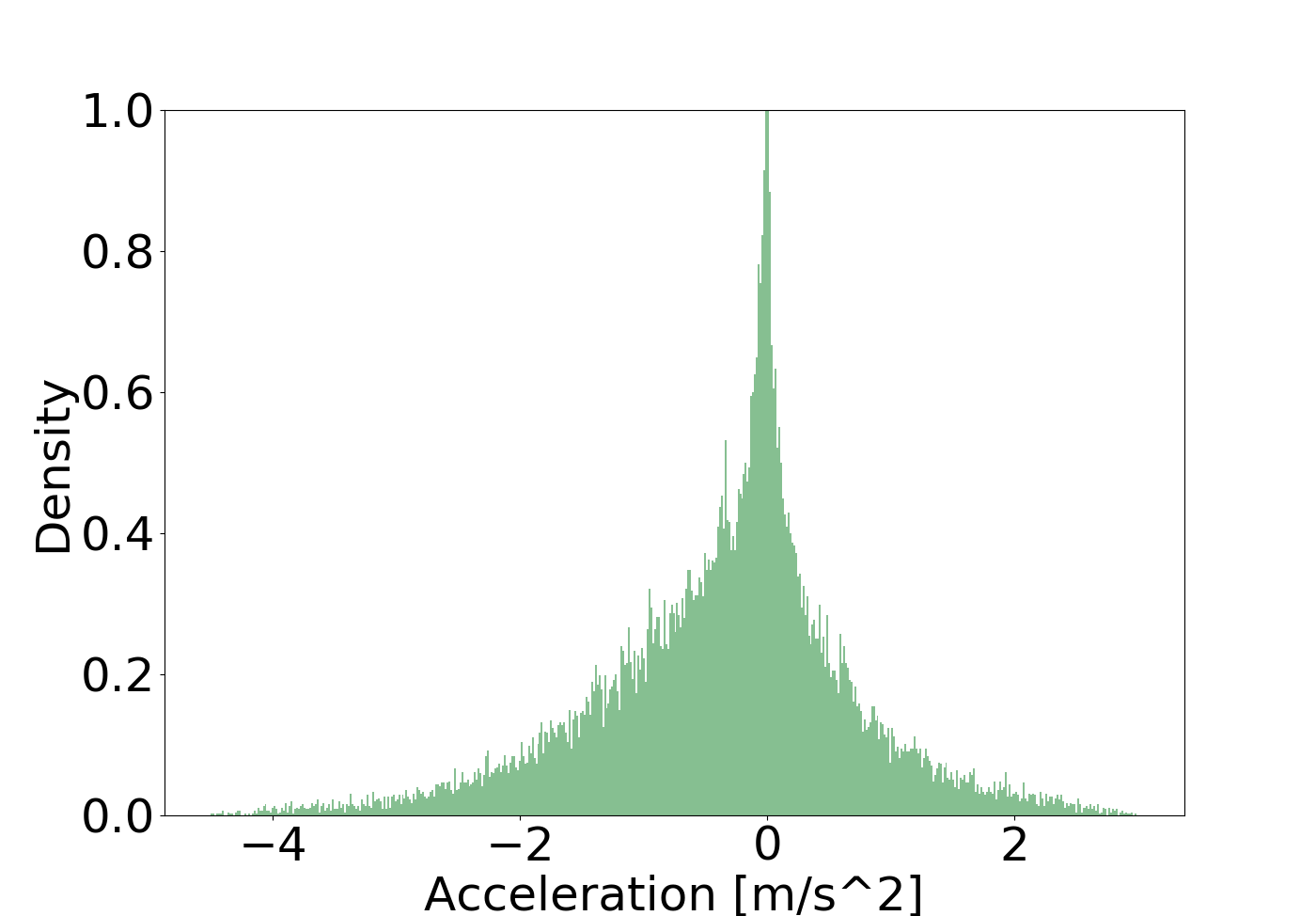}
        \caption{Ramp}
        \label{fig:acc_ramp}
    \end{subfigure}
    \hfill
    \begin{subfigure}[h!]{0.325\textwidth}
        \adjincludegraphics[width=\textwidth,Clip={0\width} {0\height} {0.075\width} {0\height}]{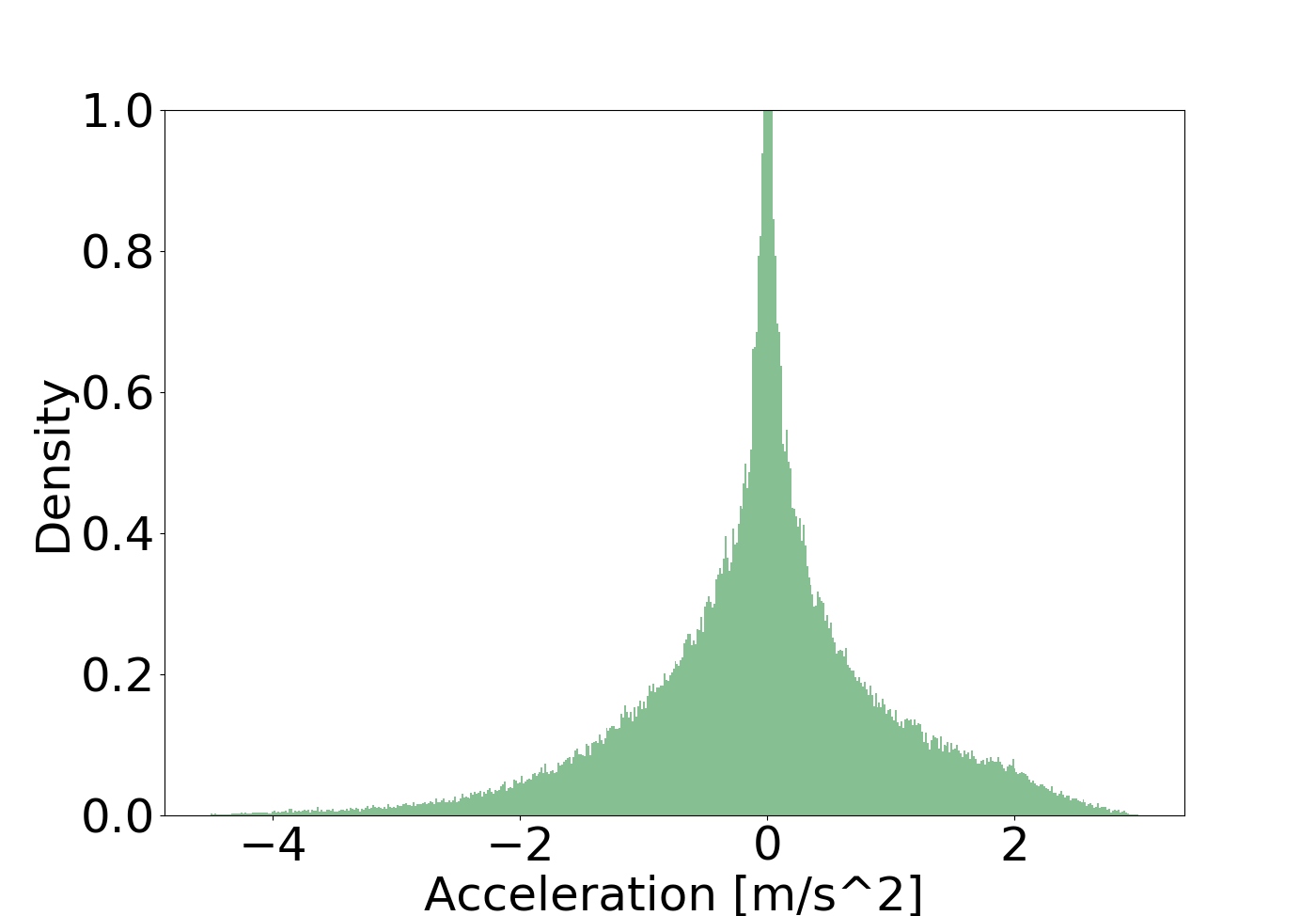}
        \caption{Both Lanes}
        \label{fig:acc_both}
    \end{subfigure}
\caption{NGSIM I-80 Acceleration Distributions}
\label{fig:acc_dist}
\end{figure}

\begin{figure}[h!]
\centering
    \begin{subfigure}[h!]{0.325\textwidth}
        \adjincludegraphics[width=\textwidth,Clip={0\width} {0\height} {0.075\width} {0\height}]{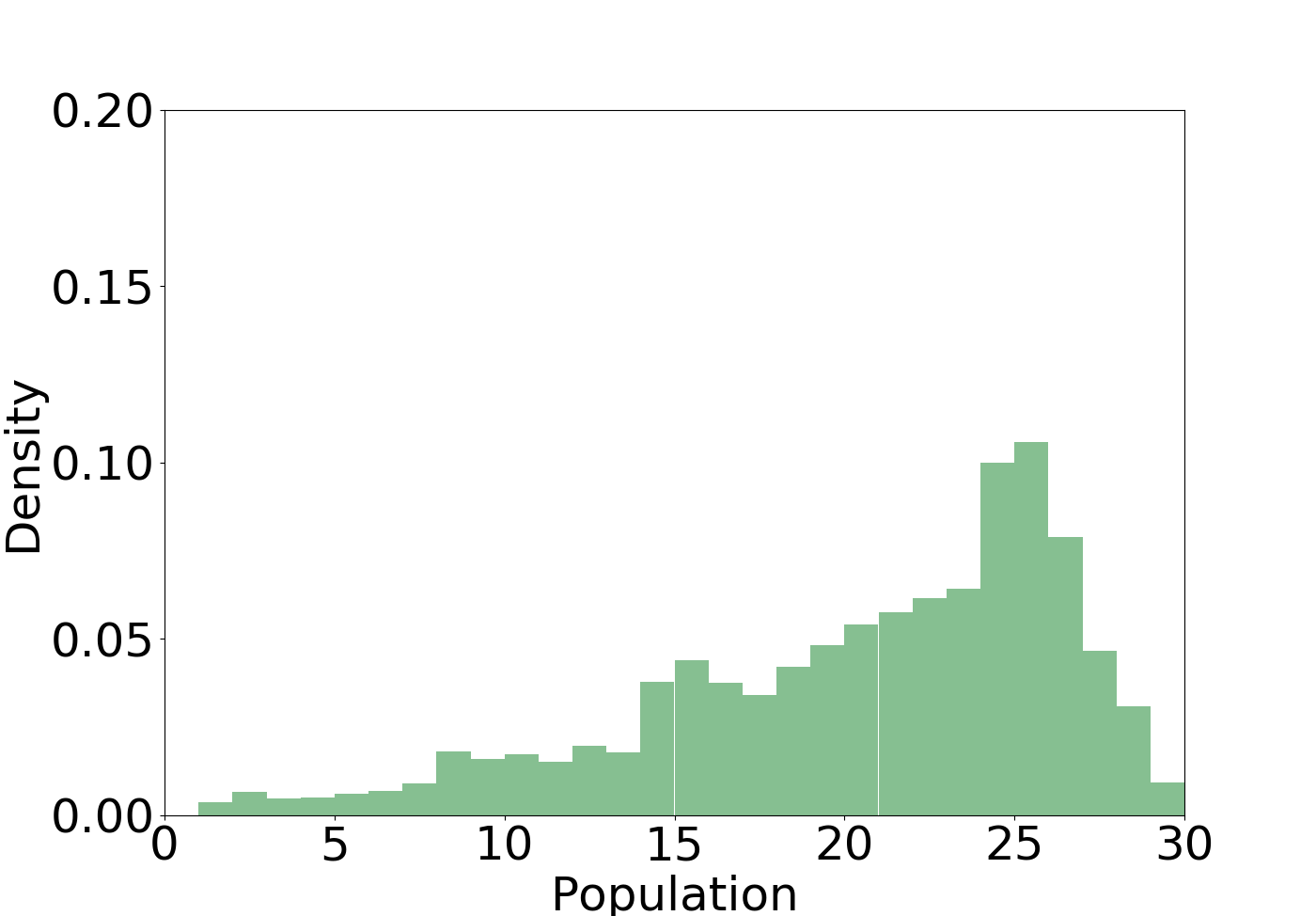}
        \caption{Main Road}
        \label{fig:pop_main}
    \end{subfigure}
    \hfill
    \begin{subfigure}[h!]{0.325\textwidth}
        \adjincludegraphics[width=\textwidth,Clip={0\width} {0\height} {0.075\width} {0\height}]{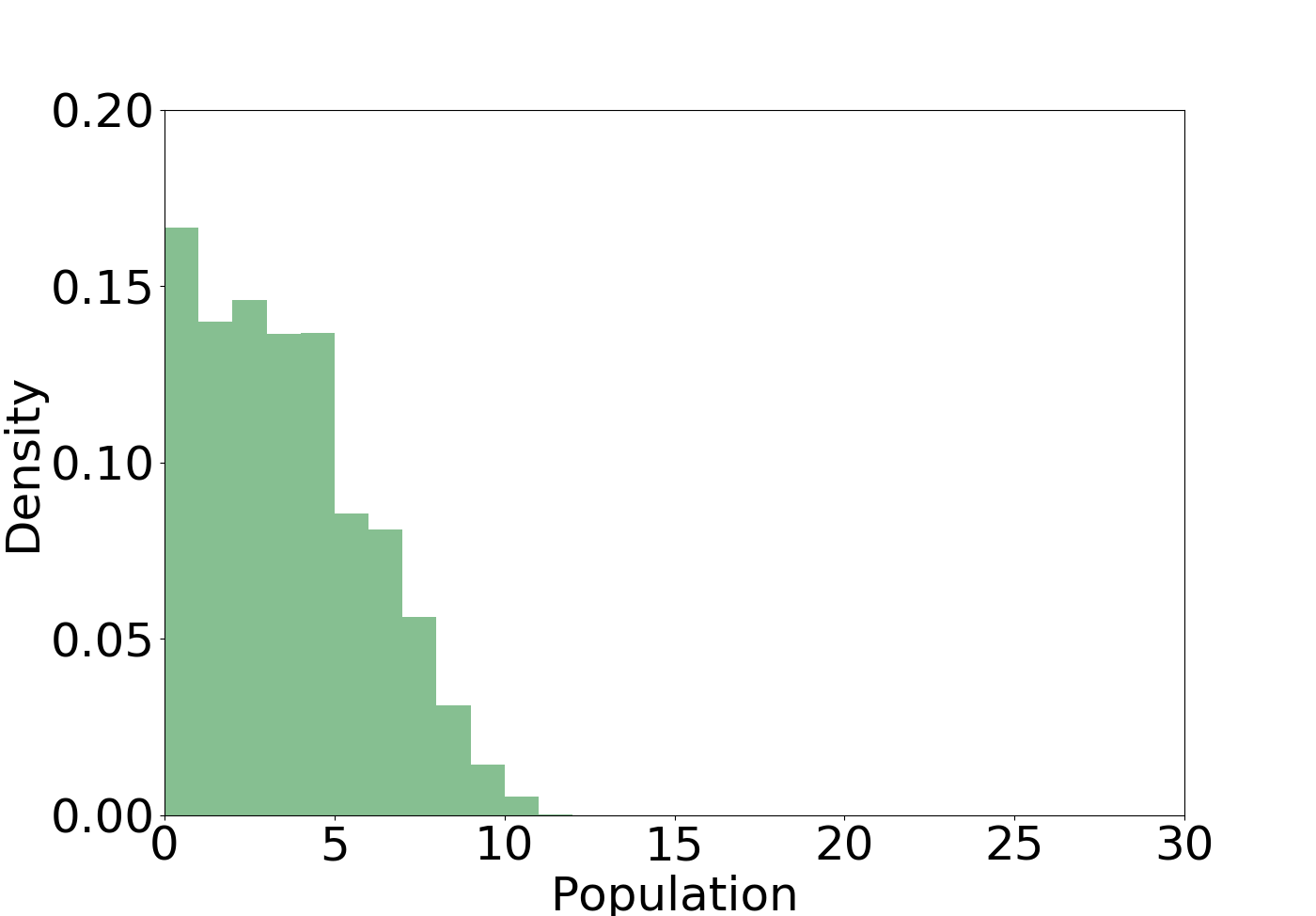}
        \caption{Ramp}
        \label{fig:pop_ramp}
    \end{subfigure}
    \hfill
    \begin{subfigure}[h!]{0.325\textwidth}
        \adjincludegraphics[width=\textwidth,Clip={0\width} {0\height} {0.075\width} {0\height}]{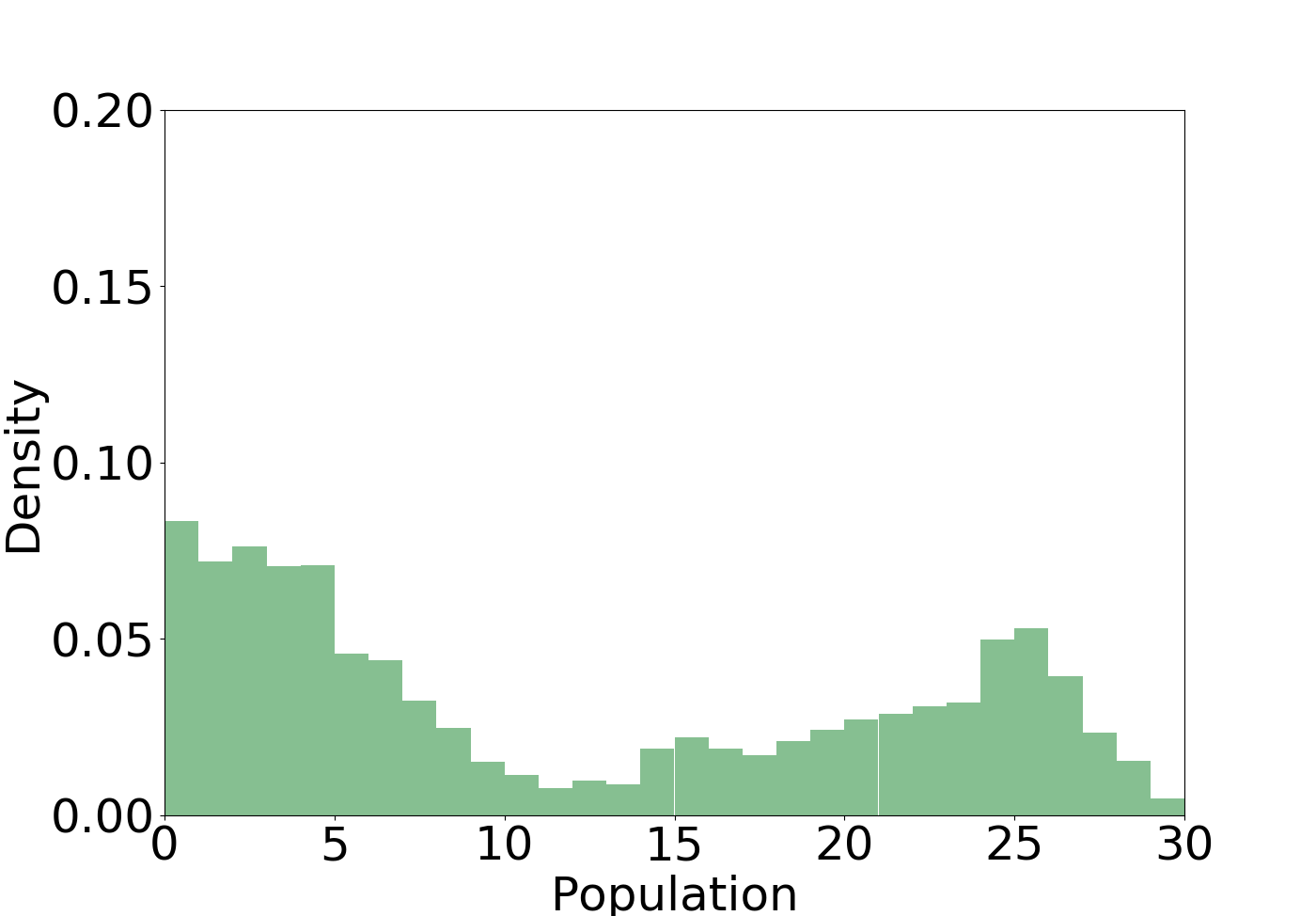}
        \caption{Both Lanes}
        \label{fig:pop_both}
    \end{subfigure}
\caption{NGSIM I-80 Population Distributions}
\label{fig:pop_dist}
\end{figure}

The dataset comprises vehicle trajectories collected from observations made for 45 minutes in total. There are three distinct 15-min periods: 4:00 p.m. to 4:15 p.m., 5:00 p.m. to 5:15 p.m., and 5:15 p.m. to 5:30 p.m. Therefore, the dataset includes congestion accumulation and the congested traffic. The sampling frequency at which the vehicle trajectories are recorded is 10 Hz. A data reconstruction of this part is needed due to excessive noise on velocity and acceleration data. \cite{MONTANINO201582} carried out such a study for the first 15 min section and we use their velocity and acceleration data for our scenario creation. Although the reconstruction provides the lane information of each vehicle, positions along the axis perpendicular to lanes are discrete, not continuous. Therefore, when a car changes its lane, the reconstructed data does not provide all the details of the transition, but simply indicates the lane change. 
Headway distance of a vehicle indicates the distance between that vehicle and the vehicle in front of it on the same lane. The distributions of headway on the main road, ramp and on both lanes are calculated using the data and shown in Fig. \ref{fig:headway_dist}. The mean and the standard deviation of the distribution on both lanes (Fig. \ref{fig:headway_both}) are 12.72 m and 10.11 m, respectively. Similarly, velocity and acceleration distributions are calculated and presented in Fig.s \ref{fig:vel_dist} and \ref{fig:acc_dist}, respectively. As expected, the vehicles traveling on the ramp commute slower than the ones on the main road. The mean and the standard deviation of the velocity distribution considering both lanes are calculated as 9.78 m/s and 4.84 m/s, respectively. Regarding the acceleration distribution, the vehicles on the ramp seem to be decelerating more frequently than the ones in the main road. Since the trend is not dominant, it is not clear whether this can be generalized to other merging scenarios. The mean and the standard deviation of the acceleration distribution on both lanes are calculated as $-0.11$  m/s$^2$ and $-0.977$ m/s$^2$, respectively. Vehicle population distributions are given in Fig. \ref{fig:pop_dist}. We observe that the main road is more congested compared to the ramp.

\FloatBarrier

Details regarding the utilization of the headway, velocity, acceleration and population distribution data for scenario generation are explained in the following sections. 

\FloatBarrier 

\FloatBarrier

\FloatBarrier

\FloatBarrier

\subsubsection{Environment Setup}

\begin{figure}[t]
    \centering
    \includegraphics[scale=.25]{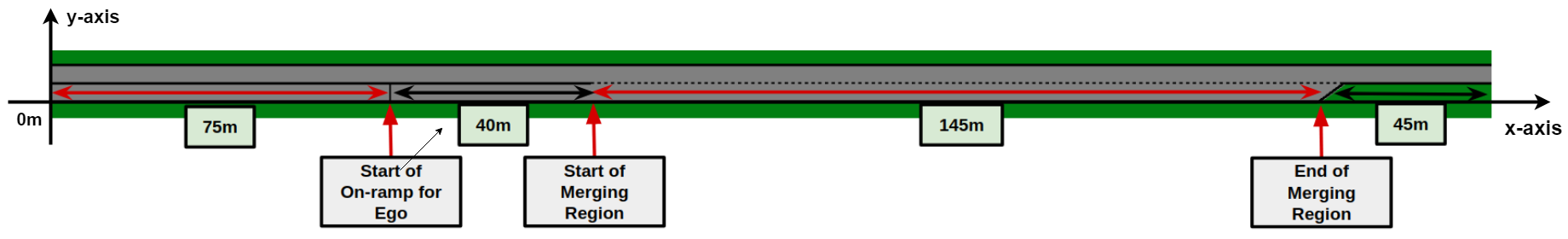}
    \caption{Highway Merging Environment}
    \label{fig:env}
\end{figure}

The environment consists of a 305 m long main road, a 185 m long ramp from \enquote*{Start of On-ramp for Ego} point and a  145 m long merging region (see Fig. \ref{fig:env}). Following the end of the merging region, the main road has an additional 45 m. The lanes in the environment have a width of 3.7 m. The vehicles are 5 m in length and 2 m in width.

The initial position of the ego vehicle is the $x=0$ m point, if placed on the main road, and $x=75$ m point, which is the \enquote*{Start of On-ramp for Ego} location, if the ego is placed on the ramp. During the initialization of the environment, vehicles are placed in a random fashion with a minimum initial distance of 2 car lengths, which is equal to 10 m. Initial velocities of the vehicles, other than the ones placed on the merging region of the ramp, are sampled from a uniform distribution on the interval [$7.78$ m/s, $11.78$ m/s], which is the $\pm$2 m/s neighborhood of the mean velocity of 9.78 m/s obtained from the I-80 dataset (see Fig. \ref{fig:vel_both}). Defining $v_{nom}=9.78$ m/s, the initial velocities of the vehicles placed on the merging region of the ramp are determined using the formula

\begin{equation}
    v(t_0) = v_{nom}*(0.5+0.5\frac{x_{e_m}-x(t_0)}{x_{e_m}-x_{s_m}})+z,
    \label{eq:initial_velocity_onramp}
\end{equation}
where $x_{s_m}$ and $x_{e_m}$ are the start and end positions of the merging region (see Fig. \ref{fig:env}), $x(t_0)$ is the initial position of the vehicle, and  $z$ is a realization of the random variable Z, which is sampled from a uniform distribution $\mathcal{U}(-2,2)$. It is noted that the closest initial position to the end of the merging region is taken to be 23 m away from the end, to allow for a safe slow down. This distance is approximately one standard deviation longer than the mean headway distance, calculated from the I-80 dataset (see Fig. \ref{fig:headway_both}). 

\FloatBarrier

The maximum number of cars that can be on the ramp is set as 7, which prevents unreasonably congested situations on the ramp. This is supported by the data as more than 95\% of the cases include 7 or fewer vehicles on the ramp (see Fig. \ref{fig:pop_ramp}). As the traffic flows, whenever a car leaves the environment, another one is added with a probability of 70\%, and its placement is selected to be the main lane with a probability of 70\%. The initial velocity of each newly added vehicle is sampled using the methods explained above. In order to add a new car safely, the minimum initial distance condition is satisfied.

\subsubsection{Agent Observation Space}
\label{sect:obs_space}

\begin{figure}[tbh]
\centering
    \begin{subfigure}[h!]{0.4\textwidth}
        \includegraphics[width=\textwidth]{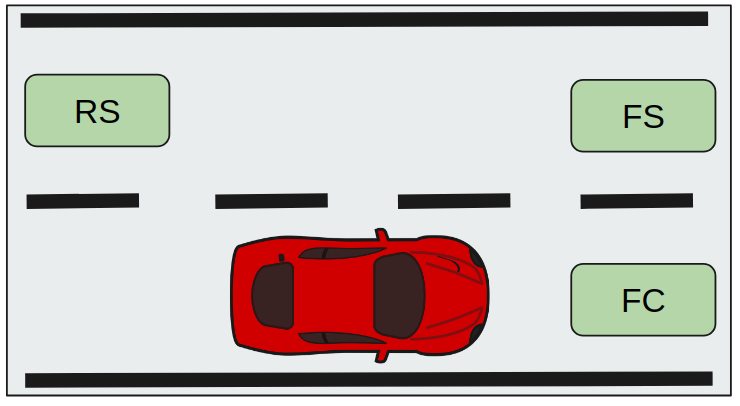}
        \caption{Ego vehicle on the ramp}
        \label{fig:obs_ramp}
    \end{subfigure}
    \hfill
    \begin{subfigure}[h!]{0.4\textwidth}
        \includegraphics[width=\textwidth]{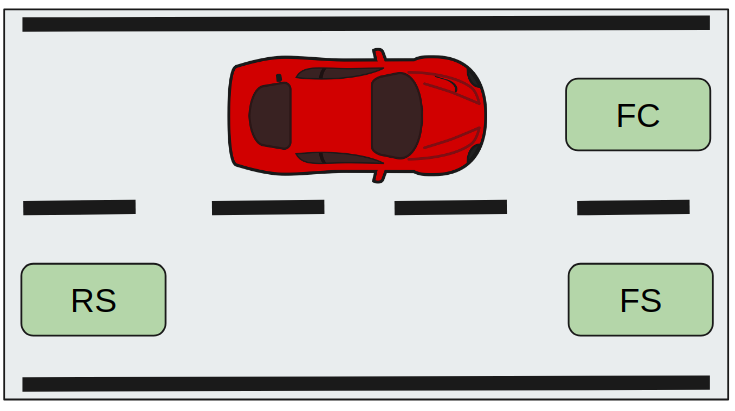}
        \caption{Ego vehicle on the main road}
        \label{fig:obs_mainroad}
    \end{subfigure}
\caption{Observation Space. RS, FS and FC refer to \textit{rear side}, \textit{front side}, and \textit{front center}, respectively.}
\label{fig:obs}
\end{figure}

In a merging scenario, a driver observes the car in front, the cars on the adjacent lane and the distance to the end of the merging region, which helps to determine when to merge or to yield the right of way to a merging vehicle.The regions used to form the observation space variables are described in Fig. \ref{fig:obs}, and the variables are given in Table \ref{table:obs_space}. 

\FloatBarrier

\begin{table} [tbh]
\centering
\caption{Observation Space Variables}
\begin{tabular}{c  c }
\hline
\textbf{Observation Space Variable} & \textbf{Normalized Range} \\
\hline
$FC_v$ &  $[-1,1]$ \\
$FC_d$ &  $[0,1]$ \\
$FS_v$ &  $[-1,1]$ \\
$FS_d$ &  $[0,1]$ \\
$RS_v$ &  $[-1,1]$ \\
$RS_d$ &  $[0,1]$ \\
$d_e$ & $[-1,1]$\\
$v_x$ & $[0,1]$\\ 
$l$ & $l \in \{0,1\}$\\
\hline
\end{tabular}
\label{table:obs_space}
\end{table}

The variables $FC$, $FS$ and $RS$ in Table \ref{table:obs_space} refer to front-center, front-side and rear-side, respectively (see Fig. \ref{fig:obs}). The subscripts $v$ and $d$ indicate whether the variable denotes relative velocity or relative distance. The normalization constants used for the relative distance and relative velocity variables are $v_{max}=29.16$ m/s and $d_{far}=23$ m, respectively. The ego driver can also observe the distance, $d_e$, to the end of the merging region. $d_e$ is normalized by the total length of the merging region, which is 45 m. Furthermore, ego can observe his/her velocity along the x-axis, $v_x$, and his/her own lane, $l$. $v_x$ is normalized with $v_{max}$, and $l$ takes the values of 0 (ramp) and 1 (main lane).  It is noted that except the lane number, all the observation space variables are continuous. 

\subsubsection{Agent Action Space}

In the highway merging problem, an agent on the main road can change only its velocity, whereas an agent on the ramp can merge, as well as change its velocity. The specific actions are given below.

\begin{itemize}
    \item $\textbf{Maintain}$: Acceleration is sampled from Laplace($\mu = 0,b = 0.1$) in the interval [$-0.25$ m/s$^2$, $0.25$ m/s$^2$]. 
    \item $\textbf{Accelerate}$: Acceleration is sampled from Exponential($\lambda = 0.75$) with starting location 0.25 m/s$^2$ and largest acceleration 2 m/s$^2$.
    \item $\textbf{Decelerate}$: Acceleration is sampled from inverse Exponential($\lambda = 0.75$) with starting location -0.25 m/s$^2$ and smallest acceleration -2 m/s$^2$.
    \item $\textbf{Hard-Accelerate}$ Acceleration is sampled from Exponential($\lambda = 0.75$) with starting location 2 m/s$^2$ and largest acceleration 3 m/s$^2$.
    \item $\textbf{Hard-Decelerate}$ : Acceleration is sampled from inverse Exponential($\lambda = 0.75$) with starting location -2 m/s$^2$ and smallest acceleration -4.5 m/s$^2$.
    \item $\textbf{Merge}$: Merging is assumed to be completed in one time-step.
\end{itemize}

\subsubsection{Vehicle Model}

Let $x(t)$, $v_x(t)$ and $a(t)$ be the position, velocity and acceleration of a vehicle along the x-axis (see Fig. \ref{fig:env}), respectively.Then, the equations of motion of the vehicle is given as

\begin{equation}
        x (t + \Delta t) = x(t) + v_x(t)* \Delta t  + 1/2 a(t)*\Delta t^2,
\end{equation}
\begin{equation}
        v_x(t + \Delta t) = v_x(t) + a(t)* \Delta t,
\end{equation}
where $\Delta t$ is the time-step, which is determined as 0.5 seconds.

\subsubsection{Reward Function}
\label{sect:Rew_fcn}

Reward function, R, is a mathematical representation of driver preferences. In this work, we use

\begin{equation}
    R = c*w_1 + h*w_2 + m*w_3 + e*w_4 + nm*w_5 + s*w_6,
    \label{eq:reward}
\end{equation}
where $w_i$'s, $i\in\{1,2,3,4,5,6\}$, are the weights used to emphasize the relative importance of each term. The terms used to form the reward function in (\ref{eq:reward}) are explained below

\begin{itemize}
    \item $c$: Collision parameter. Takes the values of -1 when a collision occurs, and zero otherwise. There are three different possible collision cases, which can be listed as
    \begin{enumerate}
        \item Ego fails to merge, and crashes into the barrier at the end of the merging region
        \item Ego merges into an environment car on the main lane
        \item Ego crashes into a car in front
    \end{enumerate}
    \item $h$: Headway distance parameter. Defining $d_{close}=3$ m, $d_{far}=13$ m and $d_{nom}=23$ m, this parameter is calculated as
    \begin{equation}
        h = 
        \begin{cases}
            -1 ,& \text{if} \hspace{1mm} FC_d \leq d_{close}\\
            \frac{FC_d - d_{nom}}{d_{nom} - d_{close}},              &\text{if} \hspace{1mm} d_{close} \leq FC_d  \leq d_{nom} \\
            \frac{FC_d - d_{nom}}{d_{far} - d_{nom}},              &\text{if} \hspace{1mm} d_{nom} \leq FC_d \leq d_{far} \\
            1,              &\text{otherwise}
        \end{cases}
        \label{eq:headway}
    \end{equation}
    The parameters $d_{close}$, $d_{far}$ and $d_{nom}$ are defined using the mean and standard deviation information obtained from the headway distribution presented in Fig. \ref{fig:headway_both}. The variable $FC_d$ is introduced in Table \ref{table:obs_space}.
    \item $m$: Velocity parameter. Calculated as
    \begin{equation}
        m = 
        \begin{cases}
            \frac{v-v_{nom}}{v_{nom}},& \text{if} \hspace{1mm} v \leq v_{nom}\\
            \frac{v_{max}-v}{v_{max}-v_{nom}},              & \text{otherwise},
        \end{cases}
    \end{equation}
    where $v$ is the velocity of the ego vehicle, $v_{nom}=9.78$ m/s corresponds to the mean of the velocity distribution in Fig. \ref{fig:vel_both}, and $v_{max}=29.16$ m/s is the speed limit on I-80. The parameter $m$ encourages the ego vehicle increase its speed until the maximum velocity is reached. 
    \item $e$: Effort parameter. Equals to zero when velocity is less than $v_{nom}/2$, and otherwise determined as
    \begin{equation}
        e = 
        \begin{cases}
            -0.25,& \text{if }\hspace{1mm} act=\textbf{Accelerate}\hspace{1mm} \text{or}\hspace{1mm} \textbf{Decelerate} \\
            -1,& \text{if }\hspace{1mm} act=\textbf{Hard-Accelerate}\hspace{1mm} \text{or}\hspace{1mm} \textbf{Hard-Accelerate} \\
            0,& \text{otherwise}
        \end{cases}
    \end{equation}
    where $act$ notes the action taken by the agent. The main purpose of the effort parameter is to discourage extreme actions such as \textbf{Hard-Accelerate} and \textbf{Hard-Decelerate} as much as possible. For low velocities, it is set to zero to encourage faster speeds so that unnecessary congestion is prevented. 
    \item $nm$: \enquote*{Not Merging} parameter. Equals to -1 when the agent is on the ramp. This parameter discourages the ego agent to keep driving on the ramp unnecessarily, without merging.   
    \item $s$: Stopping parameter. Utilized to discourage the agent from making unnecessary stops. Using the observation space variables defined in Table \ref{table:obs_space}, and the definitions of $d_{close}$ and $d_{far}$ before (\ref{eq:headway}), this parameter is defined as in the following.  
    
    If $l=1$ (main road), then 
    \begin{equation}
    s = 
    \begin{cases}
        -1,& \text{if } act\neq\textbf{Hard-Accelerate}\hspace{1mm} \text{and} \hspace{1mm} FC_d \geq d_{far}\hspace{1mm} \text{and} \hspace{1mm} d_e \geq d_{far}\\    
        0,& \text{otherwise,}
    \end{cases}
    \end{equation}    
    If $l=0$ (ramp), then 
    \begin{equation}
    s = 
    \begin{cases}
        -1,& \text{if } act\neq\textbf{Merge} \hspace{1mm} \text{and} \hspace{1mm} FS_d \geq d_{close}\hspace{1mm}  \text{and} \hspace{1mm} RS_d \leq 1.5d_{far}  \\
        -0.05,& \text{else if }\hspace{1mm} d_e \leq d_{far} \\
        0,& \text{otherwise,}
    \end{cases}
\end{equation}
\end{itemize}

\section{Training and Simulation}
\label{section:training_and_simulation}

The DQN training using dynamic level-\textit{k} game theory is carried out as explained in Algorithm \ref{alg::dynamic}. The traffic environment is constructed as described in the previous section. An episode starts once the ego and the environment vehicles are initialized, and ends when the ego car leaves the environment or experiences an accident. 

\subsection{Training Configuration}

\begin{figure}[h!]
    \centering
    \includegraphics[scale=.45]{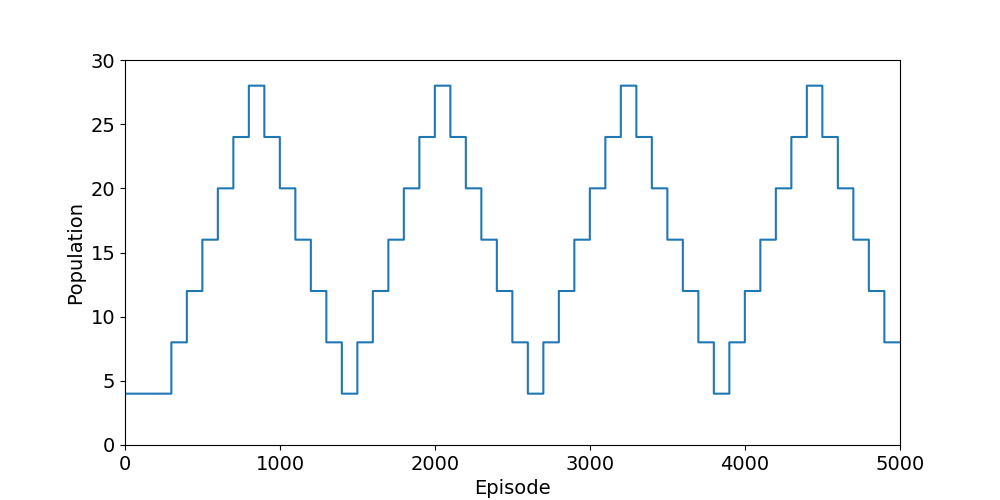}
    \caption{Population variation for Phases 1 and 2.}
    \label{fig:pop_change}
\end{figure}

Training consists of 3 different phases: Initial Phase, Sinusoidal Car Population Phase, Random Traffic Population Phase.
\begin{enumerate}
    \item \textbf{Initial Phase}: For the first 200 episodes, car population, including the ego vehicle, is set to 4. The reasoning behind first phase is to enable the agent to learn the basic dynamics of the environment under simple circumstances.
    \item \textbf{Sinusoidal Car Population Phase}: Starting at the $200^{th}$ episode until the $5000^{th}$, in every 100 episodes vehicle population is modified in a sinusoidal fashion, where the population take values from the set $\{4,8,12,16,20,24,28\}$ (see Fig. \ref{fig:pop_change}). The reasoning behind this variation is to prevent the learning agent from getting stuck at congested traffic, and to provide an environment that changes in a controlled manner.
    \item \textbf{Random Traffic Population Phase}: The car population of last 1000 episodes are altered randomly using the numbers from the set $\{4,8,12,16,20,24,28\}$, in order to expose the ego agent to a wide variety of scenarios, without letting the learning process to overfit to any scenario variation structure.  
\end{enumerate}

\FloatBarrier

\subsection{Implementation Details and Computation Specifications}

DQN for the level-\textit{k} agents is implemented via TensorFlow Keras API \citep{tensorflow2015-whitepaper}, with an  architecture that is based on a 4-layered neural network, where each layer is fully-connected. The input layer is a vector of 9 units, consisting of variables given in Table \ref{table:obs_space}. The first and second fully-connected layers are 256 units in size. The size of the third layer decreases to 128 units, and it is connected to the output layer which is comprised of 5 units that correspond to the Q-values of 5 possible actions.

DQN for the dynamic level-\textit{k} agent has the same structure, except the output layer. The output layer is comprised of 3 units that correspond to level-1, level-2 and level-3 reasoning actions.

The computation specifications of the computer used for training and simulation are

\begin{itemize}
    \item Processor: Intel\textregistered \hspace{1mm} Core\texttrademark \hspace{1mm} i7-4700HQ \@ 2.40GHz 2.40GHz
    \item RAM: DDR3L 1600MHz SDRAM, size: 16 GB.
\end{itemize}

The trainings for each level and the dynamic level-\textit{k} last for varying durations, as episodes are not time-restricted. Level-1 training lasts around 1.5 hours, whereas level-2, 3 and dynamic agents are trained in approximately 3 hours. Longer training times for higher levels and the dynamic level can be explained by the increased sophistication of the policies defined by these levels. Further details regarding the training of a DQN agent can be found in Table \ref{table:training}.

\begin{table}[tbh]
\centering
\caption{DQN Training Configuration}
\begin{tabular}{c  c}
\hline
\textbf{Hyper-parameter} & \textbf{Value} \\
\hline
Replay Memory Size & 50000 \\
Replay Start Size & 5000 \\
Target Network Update Frequency & 1000 \\
Initial Boltzmann Temperature & 50 \\
Mini-batch Size & 32 \\
Discount Factor & 0.95 \\
Learning Rate & 0.0013\\
Optimizer & Adam \\
\hline
\end{tabular}
\label{table:training}
\end{table}

\subsection{Level-0 Policy}

Level-0 policy is the anchoring piece of level-\textit{k} game theory based behavior modeling. Several approaches can be considered for designing a Level-0 agent, considering the fact that the decision-maker is non-strategic, and the decision making logic can be stochastic. The policy can be a uniform random selection mechanism \citep{SHAPIRO2014308}, or can take a single action regardless of the observation \citep{doi:10.2514/1.G000426,doi:10.2514/6.2016-1001,doi:10.2514/1.G000176}, or it can be  a conditional logic based on experience \citep{6665110}. Our level-0 policy approach is stochastic for the merging action, and rule-based for the rest of the actions. This leads to a simple but sufficiently rich set of actions during an episode, which provides enough excitation for the training of higher levels. 

Specifically, we define two level-0 policies, one for the case when the ego vehicle is on the ramp and the other for the main road. To describe these policies, we first introduce a function that provides a metric for an agent's proximity to the end of the merging region as 

\begin{equation}
    f(d_{e}(t)) = (l_m - d_e / l_m)^2,
    \label{eq:f_d_e}
\end{equation}

\noindent where $d_e$ is defined in Table \ref{table:obs_space} and $l_m$ is the length of the merging region. A quadratic function is used to reflect the rapidly increasing danger of approaching the end point. Furthermore, we define two threshold parameters as
\begin{itemize}
    \item $TTC_{hd} = 4$ s (seconds), which is the time-to-collision ($TTC$) threshold for the $\textbf{Hard-Decelerate}$ action, and
    \item $TTC_{d} = 7$ s, which is the $TTC$ threshold for the $\textbf{Decelerate}$ action.
\end{itemize}
    
\noindent We also define $\epsilon = 0.01$ to prevent division by zero cases.
Algorithms \ref{alg::level0_ramp} and \ref{alg::level0_mainroad} describe the level-0 policies for driving on the ramp and on the main road, respectively. The parameters used in these algorithms are defined in Table \ref{table:obs_space} and before (\ref{eq:headway}). 

\begin{algorithm}[H]
\caption{Level-0 Policy: On-ramp}
    \begin{algorithmic}[1]
        \State  $act$ := \textbf{Maintain}
        \If{Vehicle is in the merging region}
            
            \State Let Z be a random variable such that Z $\sim \mathcal{U}(0,1)$
            \If {$z < f(d_e)$ \textbf{or} $d_e < d_{far}$}
                
                \If {$FS_v > 0$}
                    \State $FS_v := \epsilon$
                \Else  
                 \State $FS_v := \max(-FS_v,\epsilon)$
                \EndIf
                
                \If {($\frac{FS_d}{FS_v} \geq TTC_{hd}$ \textbf{and} $FS_d > d_{close}$) \textbf{or} $FS_d > d_{far}$}
                    
                    \If {$RS_v < 0$}
                        \State $RS_v := -\epsilon$
                    \Else 
                        \State $RS_v := \max(RS_v,\epsilon)$
                    \EndIf
                    
                    \If {($\frac{-RS_d}{RS_v} \geq TTC_{hd}$ \textbf{and} $-RS_d > d_{close}$) \textbf{or} $-RS_d > 1.5d_{far}$}
                        \State $act$ := \textbf{Merge}
                    \EndIf
                
                \EndIf    
            \EndIf
        \EndIf

        \If {$act$ = \textbf{Maintain}}
            \State $FC_v := \min(FC_v,-\epsilon)$
            \State $v := v_n * d_e / l_m$
            
            \If{$(\frac{-FC_d}{FC_v} \leq TTC_{hd}$ \textbf{and} $FC_d > d_c)$ \textbf{or} $FC_d \leq d_{close}$}
                \State $act$ := \textbf{Hard-Decelerate}
            
            \ElsIf {$(\frac{-FC_d}{FC_v} \leq TTC_{d}$ \textbf{and} $FC_d > d_{close})$ \textbf{or} ($d_e < 1 0$ \textbf{and} $v > v$)}
                \State $act$ := \textbf{Decelerate}
            
            \ElsIf{$d_e \geq d_{far}$ \textbf{and} $FC_d > d_{close}$ and $FC_v > \epsilon$}
                \State $act$ := \textbf{Accelerate}
            \EndIf
        
        \EndIf
    \end{algorithmic}
    \label{alg::level0_ramp}
\end{algorithm}

\begin{algorithm}[H]
\caption{Level-0 Policy: Main road}
    \begin{algorithmic}[1]
        \State  $act$ := \textbf{Maintain}
        \State $FC_v := \min(FC_v,-\epsilon)$
        \State $v := v_{nom} * d_e / l_m$
        
        \If{$(\frac{-FC_d}{FC_v} \leq TTC_{hd}$ \textbf{and} $FC_d > d_{close})$ \textbf{or} $FC_d \leq d_{close}$}
            \State $act$ := \textbf{Hard-Decelerate}
        
        \ElsIf {$\frac{-FC_d}{FC_v} \leq TTC_{d}$ \textbf{and} $FC_d > d_{close}$}
            \State $act$ := \textbf{Decelerate}
        
        \ElsIf{$FC_d > d_{close}$ and $FC_v > \epsilon$ \textbf{and} ($v < v_{nom}$ \textbf{or} $d_e < 0$))}
            \State $act$ := \textbf{Accelerate}
        \EndIf
        
    \end{algorithmic}
    \label{alg::level0_mainroad}
\end{algorithm}

\subsection{Training Results}

\begin{figure}[h!]
\centering
    \begin{subfigure}[h!]{0.495\textwidth}
        \includegraphics[width=\textwidth]{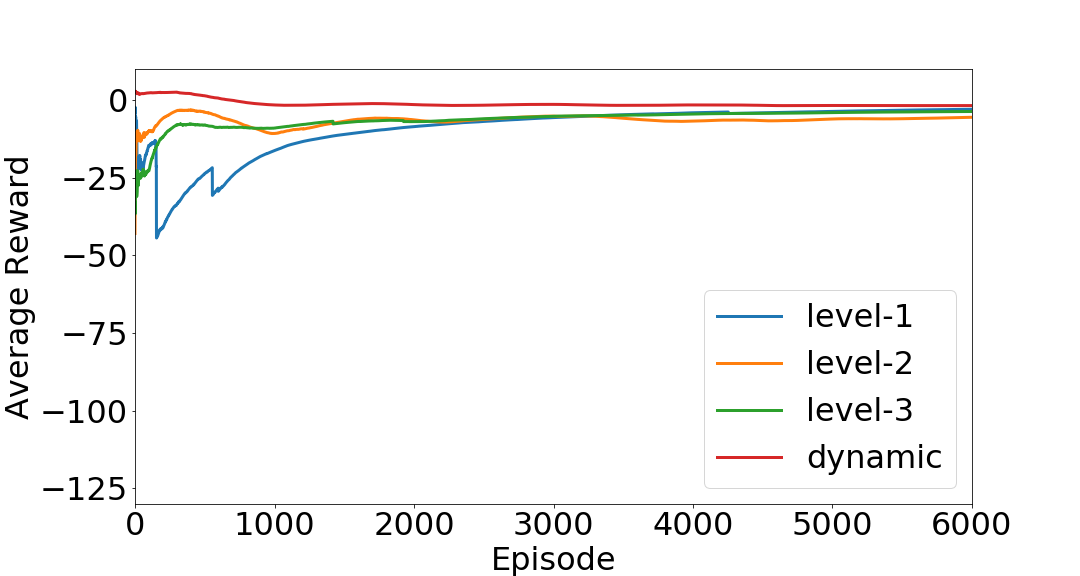}
        \caption{Average Training Reward}
        \label{fig:rewards}
    \end{subfigure}
    \hfill
    \begin{subfigure}[h!]{0.495\textwidth}
        \includegraphics[width=\textwidth]{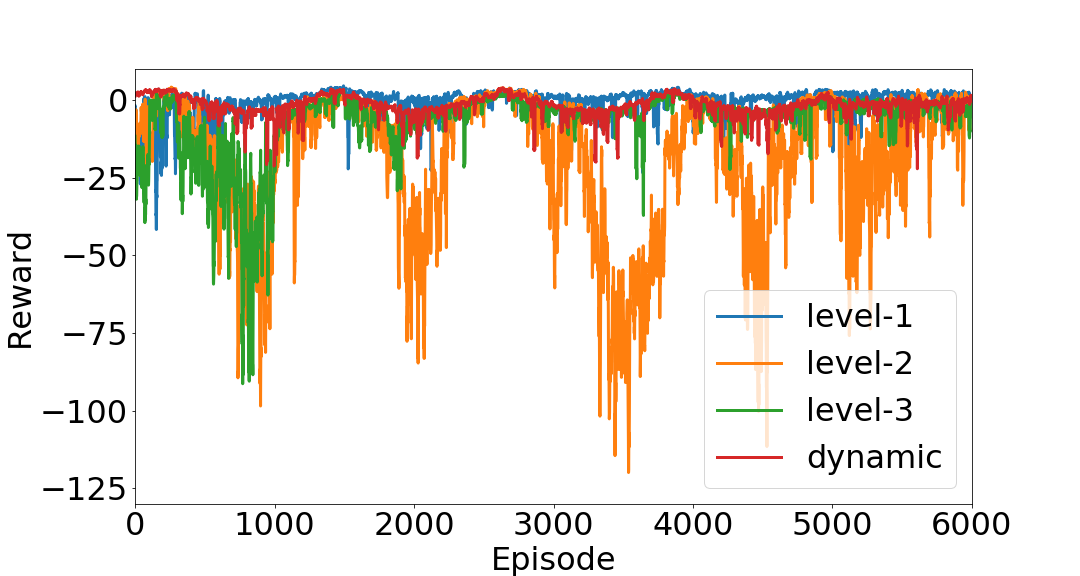}
        \caption{Reward per Training Episode}
        \label{fig:rewards2}
    \end{subfigure}
\caption{Reward Plots}
\label{fig:rewardplots}
\end{figure}

The evolutions of the average training reward and the average reward per training episode are given in Fig. \ref{fig:rewards} and \ref{fig:rewards2}, respectively. The highway merging problem brings about a commonly observed obstacle during training: bottlenecks. As the moment when the ego agent drives slowly or stops and waits for the traffic to move, it may get stuck in a local extremum and stay on the ramp or stop even though the environment traffic moves again. During training, the agent is punished for such decisions, and average reward decreases suddenly, which can be observed for level-1 training around $200^{th}$ and $900^{th}$ episodes, for level-2 training when the sinusoidal population variation arrives more crowded intervals, and for level-3 around $900^{th}$ episodes. Thanks to the stochastic nature of the exploration policy, these bottlenecks are overcome and rewards eventually increase. It is noted that at the end of the training, the last 5 models, 100 episodes apart, are simulated against themselves (level-k vs level-k), and the one with the least accidents is selected. 

\FloatBarrier

\subsection{Simulation Results}

In this section, we present simulation results that demonstrate the advantages of the new modeling framework, where the agent actions are determined through a two step sampling process: First, an appropriate reasoning level is sampled using the level selection policy trained with RL. Then, a driving action is sampled using the selected reasoning level's policy, which is also trained using RL. This is radically different than the fixed level-k method, where the actions are always sampled from a fixed policy. To investigate how the two methods compare, we tested four different ego drivers, namely the level-1, 2 and 3 drivers, and the dynamic level-\textit{k} driver. We put these drivers in different traffic settings and measure how many accidents these policies experience. The accident types are given in Section \ref{sect:Rew_fcn}. The number of vehicles in the traffic is selected from the set $\{4,8,12,16,20,24,28\}$. There are 150 episodes per population selection, which amounts to 1050 episodes in total.

Table \ref{table:sim_results} shows how four different types of ego agents perform in terms of collision rates. "Traffic Level" indicates the environment that the ego is tested in. "Level-\textit{k}" traffic consists of only level-\textit{k} agents. "Dynamic" traffic denotes an environment consisting of dynamic agents only. "Mixed" traffic, on the other hand, consists of level-0, level-1, level-2 and level-3 agents , the numbers of which are determined by uniform sampling. The table shows the drawback of a fixed level-\textit{k} policy: All level-\textit{k} policies perform significantly worse when they are placed in a level-\textit{k} traffic, compared to a level-($k-1$) traffic. This is expected since a level-k policy is trained to provide a best response to a level-($k-1$) environment, only. However, the table shows that the dynamic agent experiences the minimum collision rate in an environment consisting of dynamic agents.  Furthermore, the dynamic agent outperforms every level-\textit{k} strategy in the mixed traffic environment. These results support  the hypothesis that the agents that incorporate human adaptability and select policies that are based on human reasoning act better than static agents. 

\begin{table} [h!]
\centering
\caption{Collision rates of simulations for every ego agent type}
\begin{tabular}{@{}c c*{4}{c}}
\hline
\multicolumn{1}{c}{} & &\multicolumn{4}{c}{\textbf{Ego}}\\
\multicolumn{1}{c}{} & & Level-1 & Level-2 & Level-3 & Dynamic\\
\hline
\multirow{6}*{\rotatebox{90}{\textbf{Traffic Level}}}
& Level-0 & 1.5\% &  &  & \\ 
& Level-1 & 20.7\% & 2.3\% &  & \\
& Level-2 & & 6.9\% & 2.8\% &  \\
& Level-3 & & & 4.68\% & \\
& Dynamic & & & & 1.2\% \\
& Mixed & 37.1\% & 3.9\% & 6.1\% & 1.5\% \\
\hline
\end{tabular}
\label{table:sim_results}
\end{table}

Table \ref{table:mixed_traffic_results} shows a detailed analysis for mixed traffic results, where the number of specific types of accidents (see Section \ref{sect:Rew_fcn}) that each ego agent undergoes are presented. The numbers are normalized by 100/390, where 390 is the total number of level-1 accidents in mixed traffic for 1050 episodes. The main takeaway from this table is that dynamic agent outperforms every other level at each collision type. This explains why the dynamic agent's accident rates are lowest in a dynamic traffic environment (see Table \ref{table:sim_results}): A traffic scenario consisting of adaptable, thus more human-like, agents provides a more realistic case compared to a scenario with static agents. 

\begin{table}[h!]
\centering
\caption{Mixed traffic collision results for different levels of ego agents}
\begin{tabular}{@{}c c*{5}{c}}
\hline
\multicolumn{1}{c}{} & &\multicolumn{3}{c}{\textbf{Collision Type}}\\
\multicolumn{1}{c}{} & & Type 1 & Type 2 & Type 3 \\
\hline
\multirow{4}*{\rotatebox{90}{\textbf{Ego}}}
& Level-1 & 0 & 10.256 & 89.744 \\
& Level-2 & 0.005 & 0.054 & 0.046 \\
& Level-3 & 0.018 & 0.077 & 0.069 \\
& Dynamic & 0 & 0.008 & 0.033 \\
\hline
\end{tabular}
\label{table:mixed_traffic_results}
\end{table}
\FloatBarrier
\vspace{-.1cm}

\section{Conclusion}
\label{section:conclusion}

In this work, a dynamic level-\textit{k} reasoning, combined with DQN is proposed for driver behavior modeling. Previous studies show that level-\textit{k} game theory enables the design of realistic agents, specially in highway traffic settings when integrated with DQN. Our study builds on these early results and presents a dynamic approach which constructs agents who can decide the level of reasoning in real-time. This solves the main problem of fixed level-\textit{k} methods, where the agents do not have the ability to adapt to changing traffic conditions.  Our simulations conclude that allowing an agent to select a policy among hierarchical strategies provides a better approach while interacting with human-like agents. This paves the way for a more realistic driver modeling framework as collision rates are significantly reduced. The proposed method is also computationally feasible, and therefore naturally capable of modeling crowded scenarios, thanks to direct reasoning from observations, instead of relying on belief functions. Furthermore, the bounded rationality assumption is preserved as the available strategies are designed by adhering to the fact that humans are cognitively limited in their deductions.

\newpage
\bibliography{mybibfile}
\bibliographystyle{apa}

\end{document}